\def\year{2024}\relax
\documentclass[letterpaper]{article} % DO NOT CHANGE THIS
\usepackage{preprint_header}

\usepackage{aaai22}  % DO NOT CHANGE THIS
\usepackage{times}  % DO NOT CHANGE THIS
\usepackage{helvet}  % DO NOT CHANGE THIS
\usepackage{courier}  % DO NOT CHANGE THIS
\usepackage[hyphens]{url}  % DO NOT CHANGE THIS
\usepackage{graphicx} % DO NOT CHANGE THIS
\usepackage{natbib}  % DO NOT CHANGE THIS AND DO NOT ADD ANY OPTIONS TO IT
\usepackage{caption} % DO NOT CHANGE THIS AND DO NOT ADD ANY OPTIONS TO IT
\DeclareCaptionStyle{ruled}{labelfont=normalfont,labelsep=colon,strut=off} % DO NOT CHANGE THIS
\usepackage{algorithm}
\usepackage{algorithmic}

\usepackage{enumitem}
\usepackage{comment}
\usepackage{longtable}
\usepackage{booktabs, threeparttablex}
\setlist[itemize]{noitemsep, topsep=0pt}

\usepackage{graphicx,calc}
\newlength\myheight
\newlength\mydepth
\settototalheight\myheight{Xygp}
\newcommand*\inlinegraphics[1]{%
 \settototalheight\myheight{Xygp}%
 \settodepth\mydepth{Xygp}%
 \raisebox{-\mydepth}{\includegraphics[height=\myheight]{#1}}%
}
\usepackage{xcolor}

\usepackage{lipsum}
\usepackage{hyperref}
\usepackage{cleveref}
\usepackage{titling}
\usepackage{draftwatermark}
\SetWatermarkText{Preprint}  % Sets the watermark text
\SetWatermarkScale{1}        % Sets the scale of the watermark
\SetWatermarkColor[gray]{0.93} % Sets the color of the watermark (gray)
\SetWatermarkAngle{45}       % Sets the angle of the watermark

\usepackage{newfloat}
\usepackage{listings}
\floatstyle{ruled}
\newfloat{listing}{tb}{lst}{}
\floatname{listing}{Listing}

\title{NewsUnfold: Creating a News-Reading Application That Indicates Linguistic Media Bias  and Collects Feedback}
\author{
    Smi Hinterreiter,\textsuperscript{\rm 1}
    Martin Wessel,\textsuperscript{\rm 2}
    Fabian Schliski,\textsuperscript{\rm 3}
    Isao Echizen,\textsuperscript{\rm 4}
    Marc Erich Latoschik,\textsuperscript{\rm 1}
    Timo Spinde\textsuperscript{\rm 5}
}
\affiliations{
    \textsuperscript{\rm 1}University of Würzburg\\
    \textsuperscript{\rm 2} Technical University of Munich\\
    \textsuperscript{\rm 3}University of Passau\\
    \textsuperscript{\rm 4}National Institute of Informatics, Japan\\
    \textsuperscript{\rm 5}University of Göttingen\\
}

\begin{document}
\maketitle
\thispagestyle{preprintbox}
\begin{abstract}
Media bias is a multifaceted problem, leading to one-sided views and impacting decision-making.
A way to address digital media bias is to detect and indicate it automatically through machine-learning methods.
However, such detection is limited due to the difficulty of obtaining reliable training data.
Human-in-the-loop-based feedback mechanisms have proven an effective way to facilitate the data-gathering process.
Therefore, we introduce and test feedback mechanisms for the media bias domain, which we then implement on NewsUnfold, a news-reading web application to collect reader feedback on machine-generated bias highlights within online news articles.
Our approach augments dataset quality by significantly increasing inter-annotator agreement by 26.31\% and improving classifier performance by 2.49\%.
As the first human-in-the-loop application for media bias, the feedback mechanism shows that a user-centric approach to media bias data collection can return reliable data while being scalable and evaluated as easy to use.
NewsUnfold demonstrates that feedback mechanisms are a promising strategy to reduce data collection expenses and continuously update datasets to changes in context.
\end{abstract}

\section{Introduction} \label{sec:intro}
Media bias, slanted or one-sided media content, impacts public opinion and decision-making processes, especially on web platforms and social media \cite{effectsArd2017, oneBiasFitsAllEberl2017, SPINDE2023100264}.
News consumers are frequently unaware of the extent and influence of bias \cite{FramingClimateUncertaintyGaissmaier2019, spindeEnablingNewsConsumers2020, ribeiroMediaBiasMonitor2018}, leading to limited awareness of specific issues and narrow, one-sided points of view \cite{effectsArd2017, oneBiasFitsAllEberl2017}.
%Existing research highlights the beneficial effects of promoting media bias awareness.
As promoting media bias awareness has beneficial effects \cite{parkNewsCubeDeliveringMultiple2009, spindeHowWeRaise2022}, emphasis on the need for methods that automatically detect media bias is growing \cite{Wessel2023}.
Such methods potentially impact user behavior, as they facilitate the development of systems that analyze various subtypes of bias comprehensively and in real-time \cite{spindeYouThinkIt2021}.

Several approaches have been developed for automated media bias classification \cite{Wessel2023, spindeIntroducingMediaBias2022, liuPoliticalDepolarizationNews2021, PhrasingBiasHube2019, Spinde2023a}.
However, they share a challenge: While datasets are vital for training machine-learning models, the intricate and subjective nature of media bias makes the manual creation of these datasets time-consuming and expensive \cite{spindeNeuralMediaBias2021}.
Crowdsourcing is cost-effective but can yield unreliable annotations with low annotator agreement \cite{recasens2013a}.
In contrast, expert raters ensure consistency but lead to substantial costs \cite{spindeNeuralMediaBias2021},\footnote{For example, in the expert-based BABE dataset, one sentence label costs four to six euros, varying with rater count.} making scaling data collection challenging \cite{spindeNeuralMediaBias2021}.
Consequently, the media bias domain lacks reliable datasets for effective training of automatic detection systems \cite{Wessel2023}.
Successful Human-in-the-loop (HITL) approaches addressing similar challenges \cite{mosqueira-reyHumanintheloopMachineLearning2022, karmakharm-etal-2019-journalist} remain untested for media bias, particularly visual methods \cite{karmakharm-etal-2019-journalist}.

\begin{figure*}[!ht]
 \includegraphics[width=0.8\textwidth]{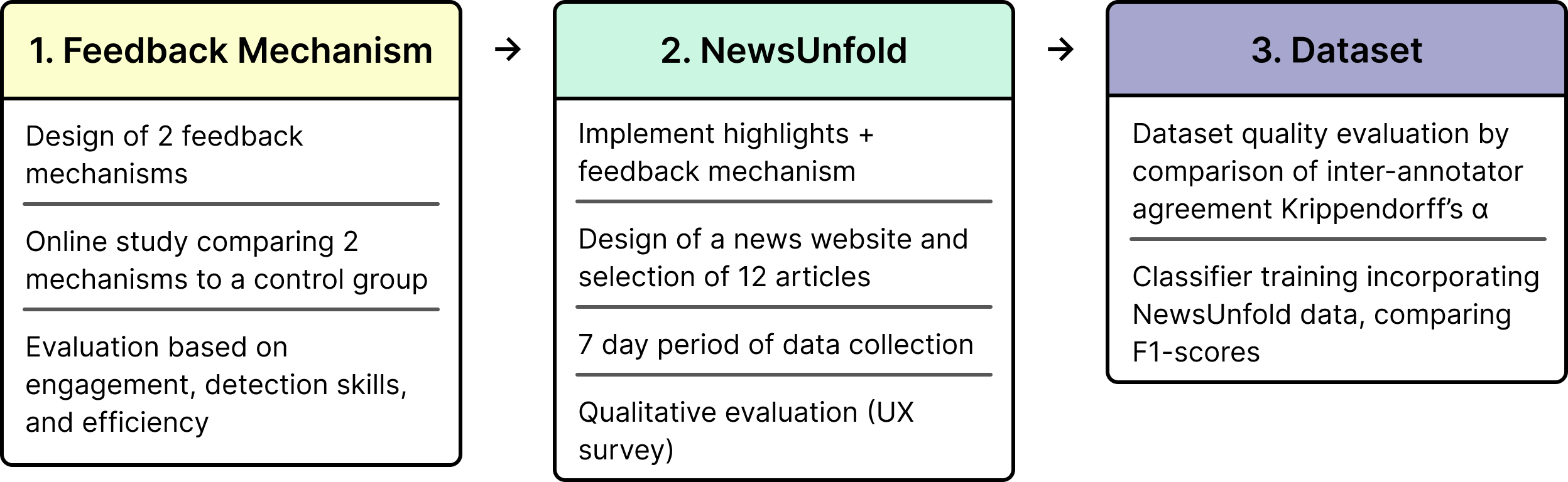}
 \centering
 \caption{Three-step process of the NewsUnfold Development and Evaluation.}
 %\Description{The three steps of this research start with designing and assessing two feedback mechanisms through a preliminary study (first tile). Both are based on an existing classifier. In the next step (second tile), the more effective feedback mechanism is integrated into the final NewsUnfold website containing 12 articles. In the last step (third tile), the data collected by the feedback mechanism is evaluated through inter-annotator agreement. Further, a classifier is trained by adding the NewsUnfold dataset to the original classifier dataset.}
 \label{fig:process}
\end{figure*}

%The study is about highlights +  feedback mechanism. The platform is an illustration of how it can be implemented in the real world
We propose a HITL feedback mechanism showcased on NewsUnfold, a news-reading platform that visually indicates linguistic bias to readers and collects user input to improve dataset quality.
NewsUnfold is the first approach employing feedback collection to gather a media bias dataset.
In the first of three phases (\Cref{fig:process}), since visual HITL (Human-in-the-Loop) methods for media bias annotation have not previously been tested, we conducted a study comparing two feedback mechanisms (\Cref{sec:feedback}).
Second, we implement a feedback mechanism on NewsUnfold (\Cref{sec:NewsUnfold}).
Third, we use NewsUnfold with 12 articles to curate the NewsUnfold Dataset (NUDA), comprising approximately 2000 annotations (\Cref{sec:studydesign}).
Notably, the collected feedback annotations exhibit a 90.97\% agreement with expert annotations and a 26.31\% higher inter-annotator agreement (IAA) than the baseline, the expert-annotated BABE dataset \cite{spindeNeuralMediaBias2021}.\footnote{The IAA evaluates how consistently different individuals assess or classify the same dataset \cite{hayesAnsweringCallStandard2007}.}
This increase is also visible when the dataset is used in classifier training, resulting in an F1-score of .824, an increase of 2.49\% compared to the baseline BABE performance.
While the platform's design is adaptable to diverse subtypes of bias, we facilitate our evaluation by focusing on linguistic bias.
Linguistic bias is defined by \citet{spindeIntroducingMediaBias2022} as a bias by word choice to transmit a perspective that manifests prejudice or favoritism towards a specific group or idea \cite{spindeIntroducingMediaBias2022}.
Despite being neither objective nor binary, collecting binary labels is a promising solution regarding the challenges arising from its ambiguous and complex nature \cite{spindeNeuralMediaBias2021}.
%added
A UX study involving 13 participants highlights high ease of use and enthusiasm for the concept.
Participants also reported a strong perceived impact on critical reading and expressed positive sentiment toward the highlights.

%We make the following contributions:
%First, we explore feedback mechanisms for the first time in the context of automated media bias detection methods.
%Second, we introduce and evaluate NewsUnfold, a news-reading platform highlighting bias in news articles, making media bias detection models accessible for everyday news consumers. NewsUnfold collects bias detection feedback and is publicly available under XXX.
%Third, we generate the NewsUnfold Dataset (NUDA) incorporating approximately 2,000 annotations. 
%Fourth, we present classifiers trained using NUDA and benchmark them against existing methodologies, enhancing performance when combined with other datasets.

In this work, we:
\begin{enumerate}
  \item Explore feedback mechanisms for the first time in the context of automated media bias detection methods.
  \item Introduce and evaluate NewsUnfold, a news-reading platform highlighting bias in news articles, making media bias detection models accessible for everyday news consumers. NewsUnfold collects feedback on bias highlights to improve its automatic detection.\footnote{All data, links, and code are publicly available at https://github.com/media-bias-group/newsunfold}
  \item Generate the NewsUnfold Dataset (NUDA) incorporating approximately 2,000 annotations.
  \item Present classifiers trained using NUDA and benchmarked against existing methodologies, enhancing performance when combined with other datasets.
\end{enumerate}

This paper proposes a design for a cost-effective HITL system to improve and scale media bias datasets.
%It is the first complete technical implementation of such a feedback system for media bias.
%That as a results of our work such feedback mechanisms offer option for all kinds of platforms
Such feedback mechanisms can be integrated into various media platforms to highlight media bias and related concepts.
Further, the system can adapt to changes in language and context, facilitating applied endeavors to run models on news sites and social media to understand and mitigate media bias and increase readers' awareness.

\section{Related Work}\label{sec:bg}
\subsection{Media Bias}\label{sec:bg:mediabias}
Various studies \cite{leeNeuSNeutralMultiNews2022, recasens2013a, razaDbiasDetectingBiases2022, PhrasingBiasHube2019, effectsArd2017, oneBiasFitsAllEberl2017} highlight the complex nature of media bias, or, more specifically, linguistic bias \cite{recasens2013a, Wessel2023, spindeIntroducingMediaBias2022}.
Individual backgrounds, such as demographics, news consumption habits, and political ideology, influence the perception of media bias \cite{impactDruckman2005, eveland2003impact, effectsArd2017, FramingClimateUncertaintyGaissmaier2019}.
Content resonating with a reader's beliefs is often viewed as neutral, while dissenting content is perceived as biased \cite{FramingClimateUncertaintyGaissmaier2019, feldmanPartisanDifferencesOpinionated2011}.
Enhancing awareness of media bias can improve the ability to detect bias at various levels --- word-level, sentence-level, article-level, or outlet-level \cite{spindeHowWeRaise2022, simpleFramingInterventionBaumer2015}.

While misinformation is closely connected to media bias and has received much research attention, most news articles do not fall into strict categories of veracity \cite{IndividualDifferencesRisk}.
Instead, they frequently exhibit varying degrees of bias, underlining the importance of media bias research.

\subsection{Automatic Media Bias Detection} \label{sec:bg:detection}
NLP methods can automate bias detection, enabling large-scale bias analysis and mitigation systems \cite{Wessel2023, spindeNeuralMediaBias2021, liuPoliticalDepolarizationNews2021, leeNeuSNeutralMultiNews2022, pryzant_automatically_2020, heDetectPerturbNeutral2021}.
Yet, current bias models' reliability for end-consumer applications is limited \cite{spindeNeuralMediaBias2021} due to their dependency on the training dataset's quality.
These models often rely on small, handcrafted, and domain-specific datasets, frequently using crowdsourcing \cite{Wessel2023}, which cost-effectively delegates annotation to a diverse, non-expert community \cite{XINTONG20147987}.
The subjective nature of bias and potential inaccuracies from non-experts can result in lower agreement, more noise \cite{spindeMBICMediaBias2021}, and the perpetuation of harmful stereotypes \cite{otterbacherCrowdsourcingStereotypesLinguistic2015}.
Conversely, expert-curated datasets offer higher agreement but come at a greater cost \cite{spindeIntroducingMediaBias2022}.
%while expert-curated datasets offer higher agreement but at a greater cost \cite{spindeIntroducingMediaBias2022}.

Datasets used for automated media bias detection need to stay updated \cite{Wessel2023}, annotations should be collected across demographics \cite{pryzant_automatically_2020}, and media bias awareness reduces misclassification \cite{spindeNeuralMediaBias2021}.
The limited range of topics and periods covered by current datasets and the complexities involved in annotating bias decreases the accuracy of media bias detection tools.
This, in turn, impedes their widespread adoption and accessibility for everyday users \cite{spindeIntroducingMediaBias2022}.
To make the data collection process less resource-intensive and optimize gathering human feedback, we raise media bias awareness by algorithmically highlighting bias and gathering feedback from readers.

\subsection{Media Bias Awareness}
News-reading websites like AllSides\footnote{\url{https://www.allsides.com/}} or GroundNews\footnote{\url{https://ground.news/}} offer approaches for media bias awareness at article and topic levels \cite{spindeHowWeRaise2022, An_Cha_Gummadi_Crowcroft_Quercia_2021, parkNewsCubeDeliveringMultiple2009}.
However, research on these approaches is sparse.
One approach uses ideological classifications \cite{An_Cha_Gummadi_Crowcroft_Quercia_2021, parkNewsCubeDeliveringMultiple2009, yaqubEffectsCredibilityIndicators2020} to show contrasting views at the article level.
At the text level, studies use visual bias indicators like bias highlights \cite{spindeEnablingNewsConsumers2020, spindeHowWeRaise2022, simpleFramingInterventionBaumer2015} with learning effects persisting post-highlight removal \cite{spindeHowWeRaise2022}.
As the creation of media bias datasets does not include media bias awareness research, NewsUnfold connects these research areas.

\subsection{HITL Platforms For Crowdsourcing Annotations}
HITL learning improves machine learning algorithms through user feedback, refining existing classifiers instead of creating new labels \cite{mosqueira-reyHumanintheloopMachineLearning2022, Sheng_Zhang_2019}.
Enhanced classifier precision can be achieved by combining crowdsourcing and HITL approaches, leveraging user feedback to generate labels via repeated-labeling, and increasing the number of annotations \cite{XINTONG20147987, karmakharm-etal-2019-journalist, Sheng_Zhang_2019, stumpf-2007-user-feedback}.
%related example 1, similar topic bc of news but experts
For instance, "Journalists-In-The-Loop" \cite{karmakharm-etal-2019-journalist} continuously refines rumor detection by soliciting visual veracity ratings from journalist's feedback.
Similarly, \citet{HITLmediaBias2018} suggest a HITL system to detect media bias in videos.
%added
They plan to extract bias cues through comparative analysis and sentiment analysis and rely on scholars to validate the output.
However, their system stays in the conceptual phase.
%related example 2, actually a platform and on news
\citet{brew-crowdsourcing-2010}'s web platform crowdsources news article sentiments and re-trains classifiers based on non-expert majority votes, emphasizing the effectiveness of diversified annotations and user demographics over mere annotator consensus.
\citet{HITLChallengesMisinformation} propose combining automatic methods, crowdsourced workers, and experts to balance cost, quality, volume, and speed.
Their concept uses automated methods to identify and classify misinformation, passing some to the crowd and experts for verification in unclear cases.
Similar to \citet{HITLmediaBias2018}, they do not implement their system and describe no UI details.
%added

%the argumentation for the comparison mechanism
As no HITL system has been implemented to address media bias, we aim to close this gap by integrating automatic bias highlights based on expert annotation data readers can review.
To mitigate possible anchoring bias and uncritical acceptance of machine judgments, we test a second feedback mechanism aimed at increasing critical thinking \cite{vaccaro2019effects, FURNHAM201135, cowritingOpinionatedLanguageModel, ShawIncentives2011}.

\section{Feedback Mechanisms}
\label{sec:feedback}
As the evaluation of feedback mechanisms for media bias remains unexplored, in a preliminary study, we design and assess two HITL feedback mechanisms for their suitability for data collection.
%Using news article sentences, labeled by the classifier from \citet{spindeIntegratedApproachDetect2020}, we compare the mechanisms \textit{Highlights}, \textit{Comparison}, and a control group that lacks visual highlights to discern (1) dataset quality using Krippendorff's $\alpha$, (2) engagement, quantified by feedback given on each sentences\footnote{Readers can modify their annotations at any time; however, each unique sentence annotation is counted as a single interaction for our feedback metric.}, (3) agreement with expert annotations through F1 scores, and (4) feedback efficiency, measured by needed time in combination with engagement and agreement.
Using sentences from news articles labeled by the classifier from \citet{spindeIntegratedApproachDetect2020}, we compare the mechanisms \textit{Highlights}, \textit{Comparison}, and a control group without visual highlights.
Our analysis focuses on (1) dataset quality, assessed using Krippendorff's $\alpha$; (2) engagement, quantified by feedback given on each sentence\footnote{Readers can modify their annotations at any time; however, each unique sentence annotation counts as a single interaction for our feedback metric.}; (3) agreement with expert annotations, evaluated through F1 scores; and (4) feedback efficiency, measured by the time required in combination with engagement and agreement.
%\begin{itemize}
%\item The dataset quality using Krippendorff's $\alpha$.
%\item Engagement, quantified by feedback-related interactions.
%\item Agreement with expert annotations through F1 scores.
%\item Feedback efficiency, measured by needed time in combination with engagement and agreement.
%\end{itemize}

\begin{figure*}[ht]
 \includegraphics[width=\textwidth]{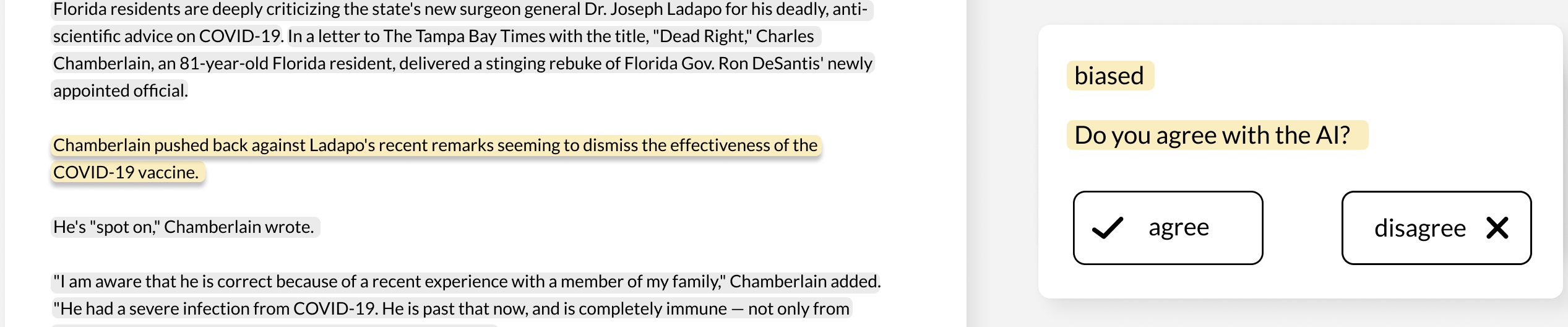}
 \caption{The feedback mechanism \textit{Highlights} uses the BABE classifier to highlight biased sentences in yellow and not biased sentences in grey. Readers can agree or disagree with this classification through the feedback module on the right.}
 %\Description{The figure shows sentences from a news article and the feedback module to its right. The feedback mechanism Highlights uses the classifier to highlight biased sentences in yellow and not biased sentences in grey. Readers can agree or disagree with this classification through the feedback module on the right. The feedback module states, "biased. Do you agree with the AI?" (highlighted in yellow). Below are an "agree" button and a "disagree" button.}
 \label{fig:highlights}
\end{figure*}

\begin{figure*}[ht]
 \includegraphics[width=\textwidth]{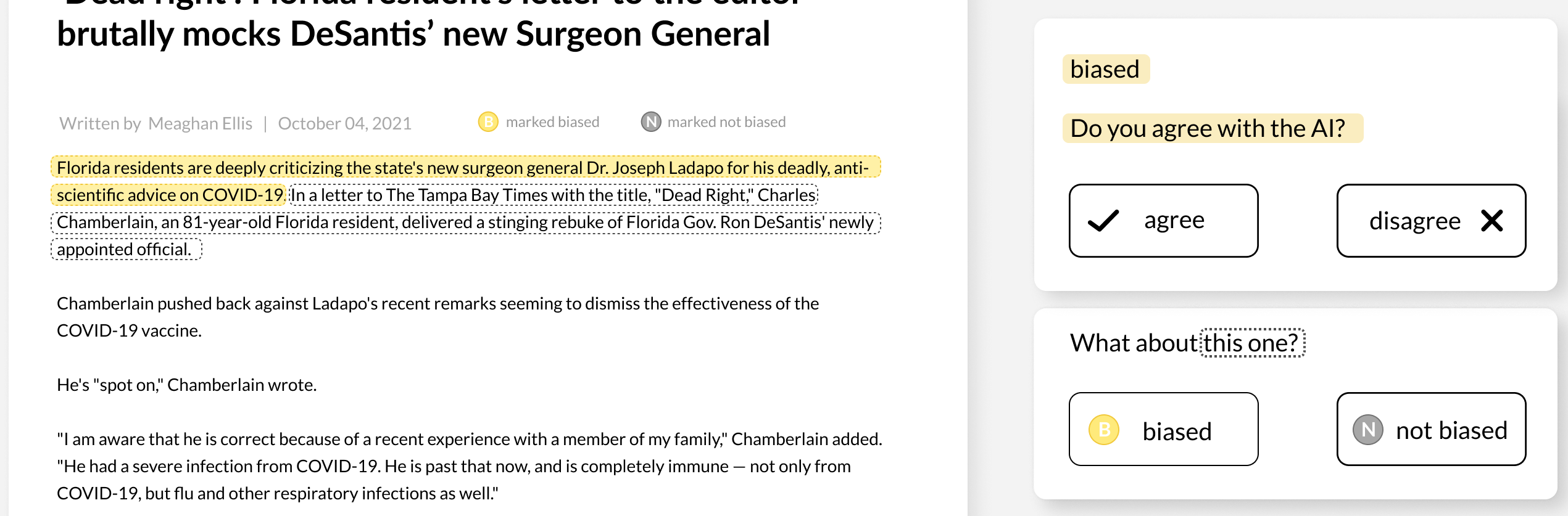}
 \caption{The feedback mechanism \textit{Comparison} operates on sentence pairs and uses the BABE classifier to highlight the first sentence as biased in yellow. Readers can agree or disagree with this classification through the feedback module on the right. The next sentence is merely outlined. Here, the feedback module asks for a bias rating without the classifier anchor.}
 %\Description{The figure shows sentences from a news article and two feedback modules to its right. The feedback mechanism Comparison highlights the first (biased) sentences in yellow. Readers can agree or disagree with this classification through the feedback module on the right. The feedback module states, "biased. Do you agree with the AI?" Below are an "agree" button and a "disagree" button. The next sentence is merely outlined. The second feedback module below the first asks, "What do you think?" (outlined like the sentence) instead. Two buttons below offer the options "biased" and "not biased." The following sentences are not highlighted or outlined.}
 \label{fig:comparison}
\end{figure*}

In the \textit{Highlights} mechanism, biased sentences are colored yellow, and non-biased ones are grey, inspired by \citet{spindeHowWeRaise2022}.
Participants indicate their agreement or disagreement with these classifications through a floating module (\Cref{fig:highlights}).
The \textit{Comparison} mechanism displays sentence pairs.
For the first sentence, participants provide feedback on the AI's classification as in \textit{Highlights}.
The second sentence has no color coding, prompting users with "What do you think?" (\Cref{fig:comparison}), thereby aiming to foster an independent bias assessment and mitigate anchoring effects.
Participants in the control group do not see any highlights, solely encountering the feedback module with the second question from \textit{Comparison}.

%#1 describe how BABE works
We use the BABE classifier trained by \citet{spindeNeuralMediaBias2021} to generate the sentence labels and highlights.
Currently, the classifier showcases the highest performance by fine-tuning the large language model RoBERTa with an extensive dataset on linguistic bias annotated by experts on both sentence and word levels.
The BABE-based model on Huggingface\footnote{https://huggingface.co/mediabiasgroup/da-roberta-babe-ft} generates the probability of a sentence being biased or not biased for each article.
We accordingly assign the label with the higher probability.

\subsection{Study Design}
To assess the two mechanisms, we recruit 240 participants, balanced regarding gender, from Prolific.\footnote{\url{https://www.prolific.co}}
On the study website built for this purpose, depicted in \Cref{fig:FeedbackHighlights}, they view two articles from different political orientations paired with one feedback mechanism per group.
During the study, users freely determine their annotation count and time spent, with a progress bar showing the number of annotated sentences.
Not interacting with any sentences prompts a pop-up, but they can click 'next' to proceed.

Curated from AllSides, articles match the baseline dataset's topics \cite{spindeNeuralMediaBias2021} and were annotated by four experts.\footnote{Experts have at least six months experience in media bias. Consensus was achieved through majority or discussion.}
\Cref{tab:FeedbackArticles} compares the classifier and expert annotations.
To measure IAA, we use Krippendorff's $\alpha$, an evaluation metric often used in the media bias domain that assesses dataset quality by determining annotator agreement beyond chance \cite{hayesAnsweringCallStandard2007}.
%#13
As higher engagement yields more data, we measure engagement through the number of decisions made with the feedback mechanism.
An efficient feedback mechanism reduces the task's tedium while ensuring data quality.
Efficiency is calculated with the Bonferroni correction: \(\frac {Engagement}{Time}*F1\).

We guarantee GDPR conformity through a preliminary data processing agreement.
A demographic survey and an introduction to media bias follow (\Cref{sec:demoSurv}).
A post-introduction attention test confirms participants' understanding of media bias, which, if failed twice, results in study exclusion.
Then, participants read through a description of the study task and proceed to give feedback on the two articles.
Lastly, a concluding trustworthiness question ensures data reliability.
%#6
If participants clicked through the study inattentively, they could indicate that their data is not usable for research \cite{drawsChecklistCombatCognitive2021} while still receiving full pay \cite{spindeHowWeRaise2022}.

%\begin{table*}[]
%\centering
%\begin{tabular}{l|c|c|c|c}
%\toprule
% & Experts biased & Experts not biased & Classifier biased & Classifier not biased\\\midrule
%Left article & 16 & 21 & 8 & 29 \\
%Right article & 24 & 21 & 12 & 33 \\\bottomrule
%\end{tabular}
%\caption{Bias rating of sentences in study articles by classifier and experts.}
%\label{tab:FeedbackArticles} 
%\end{table*}

\begin{table*}[h!]
\centering
\begin{tabular}{l|l|l|l|l|l}
\toprule
Group & Feedback$^a$ & Engagement & IAA & F1-Score & Efficiency$^b$ \\
\midrule
Highlights & 5564 &$.9329 \pm .1642$ &.229 & $.5720 \pm .1266$ & $.1252 \pm .0951$ \\
Comparison & 4484 &$.8088 \pm .3266$ &.22 & $.5736 \pm .1339$ & $.0813 \pm .0421$ \\
Control & 5037 &$.8690 \pm .2853$ &.2 & $.5769 \pm .1566$ & $.1116 \pm .0678$ \\
\bottomrule
\multicolumn{5}{l}{$^a$ Number of feedback-related interactions}\\
\multicolumn{5}{l}{$^b$ Calculated based on the Bonferroni correction}\\
\end{tabular}
\caption{Overview of Feedback Interactions per Group.}
\label{tab:FeedbackResults} 
\end{table*}

\subsection{Results}
The 240 participants in the study spent an average of 11:24 minutes, with a compensation rate of £7.89/hr.
Twelve participants failed the attention test once, but only one was excluded for a second failure.
We further excluded 33 participants who flagged their data as unsuitable for research.
% in the concluding trustworthiness query. % -> 14%
Therefore, the analysis includes data from 206 participants: 69 control group participants, 66 \textit{Comparison} group participants, and 71 \textit{Highlights} group participants ($p = .84, f = .23, \alpha = .05$).
%here demographics
104 participants identified as female, 99 as male, and 3 as other, with an average age of 36.62 years ($SD = 13.74$).
The sample, on average, exhibits a left slant (\Cref{fig:v1:pol} and \Cref{fig:v1:polMean}) with higher education (\Cref{fig:v1:education}).
196 participants indicated advanced English levels, 9 intermediate, and 1 beginner (\Cref{fig:v1:english}).
News reading frequency averaged around once a day (\Cref{fig:v1:news}).

Notably, we observe a high overall engagement, with even the least annotated sentences receiving feedback from 70\% of the participants.
We detail the results of the feedback mechanism study, including engagement, IAA, F1 scores, and efficiency, in \Cref{tab:FeedbackResults}.
The \textit{Highlights} group exhibits higher engagement than the \textit{Comparison} group, containing more collected data.
Also, \textit{Highlights} demonstrates higher efficiency by collecting more feedback data in less time without compromising quality measured by IAA and agreement with the expert standard.

The increases in engagement and efficiency are significant at a .05 significance level.
Due to variance inhomogeneity indicated by a significant Levene test (p \textless .05), we applied Welch’s ANOVA for unequal variances.
Post-hoc Holm-Bonferroni adjustments revealed significant differences between the CONTROL and HIGHLIGHTS groups, with p \textless .0167 for efficiency and p \textless .025 for engagement.
The Games-Howell post-hoc test confirmed these results.%Levene test significant (p < 0.05) -> Welch’s ANOVA -> Holm-Bonferroni adjusted: p < 0.0167, Games-Howell test due to variance inhomogeneity. COMPARISON - CONTROL and COMPARISON - HIGHLIGHTS (Holm-Bonferroni adjusted, both p < 0.0167)
As in previous research, IAA and F1 scores from crowdsourcers are low due to the complex and subjective task \cite{spindeMBICMediaBias2021}.
%added
F1 score differences are not significant (ANOVA with Holm-Bonferroni, p \textgreater .05).
Given the comparable IAA and F1 scores across groups, we integrate \textit{Highlights} within NewsUnfold to optimize data collection efficiency.
%F1 Scores: Levene test is not significant (p > 0.05) -> regular one-way ANOVA for significance -> not significant

\begin{table*}[hbt!]
\centering
\begin{tabular}{l|p{6cm}|l}
\toprule
\textbf{Element} & \textbf{Description} & \textbf{Reference} \\
\midrule
\inlinegraphics{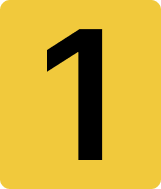} Article text with bias highlights & Presents the text with bias indicators, enhancing media bias awareness. Includes headline, author, outlet, and metadata. & \inlinegraphics{figures/icons/1.png} in \Cref{sec:feedback}, \Cref{fig:feedbackSparkles}, \Cref{fig:NewsUnfold} \\
\hline
\inlinegraphics{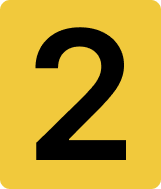} Feedback mechanism \textit{Highlights} & Integrated \textit{Highlights} mechanism, prompting users to consider the sentence and feedback on the classification. & \inlinegraphics{figures/icons/2.png} in \Cref{sec:feedback}, \Cref{fig:feedbackSparkles}, \Cref{fig:NewsUnfold} \\
\hline
\inlinegraphics{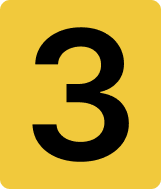} Free-text field for reasoning & Field where readers explain their bias assessments for more thorough feedback. & \inlinegraphics{figures/icons/3.png} in \Cref{sec:NewsUnfold}, \Cref{fig:feedbackSparkles}, \Cref{fig:NewsUnfold} \\
\hline
\inlinegraphics{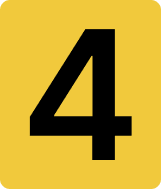} \textit{Sparkles} indicators & Emphasizing unannotated or controversial sentences, prompting further feedback. & \inlinegraphics{figures/icons/4.png} in \Cref{sec:NUdesign}, \Cref{fig:feedbackSparkles}, \Cref{fig:NewsUnfold} \\
\hline
\inlinegraphics{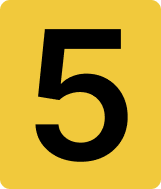} UX survey button & Prompt for a UX survey, collecting feedback on app usability and satisfaction. & \inlinegraphics{figures/icons/5.png} in \Cref{sec:NUdesign}, \Cref{fig:recommended} \\
\hline
\inlinegraphics{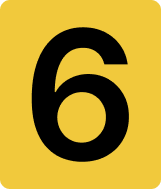} Recommended articles & Displays three suggested articles, prompting continued reading based on user behavior and article annotations. & \inlinegraphics{figures/icons/6.png} in \Cref{sec:NUdesign}, \Cref{fig:recommended} \\
\hline
\inlinegraphics{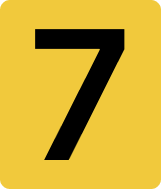} Tutorial & Gives readers a tour of NewsUnfold, explains media bias, and the feedback mechanism & \inlinegraphics{figures/icons/7.png} in \Cref{sec:NUdesign}, \Cref{fig:articlesOverview} \\
\bottomrule
\end{tabular}
\caption{Key Elements of NewsUnfold. Yellow numbers appear in NewsUnfold figures.}
\label{table:key-elements}
\end{table*}

\section{The NewsUnfold Platform} \label{sec:NewsUnfold}
%#4
Tailored toward news readers, NewsUnfold highlights potentially biased sentences in articles (\inlinegraphics{figures/icons/1.png} in \Cref{fig:feedbackSparkles}) and incorporates the \textit{Highlights} feedback module (\inlinegraphics{figures/icons/2.png} in \Cref{fig:feedbackSparkles}) assessed in \Cref{sec:feedback} to create a comprehensive, cost-effective, crowdsourced dataset through reader-feedback.
The feedback mechanism additionally includes a free-text field (\inlinegraphics{figures/icons/3.png} in \Cref{fig:feedbackSparkles}) where readers can justify their feedback.

\subsection{Application Design}\label{sec:NUdesign}
NewsUnfold's responsive design draws inspiration from news aggregation platforms,\footnote{E.g., Google News (\url{https://news.google.com}).} aiming to represent an environment where users, given updating content, return to regularly.
By clarifying the purpose of our research, the societal importance of media bias, and giving access to automated bias classification, we encourage voluntary feedback contributions.

\begin{figure*}[ht]
 \includegraphics[width=\textwidth]{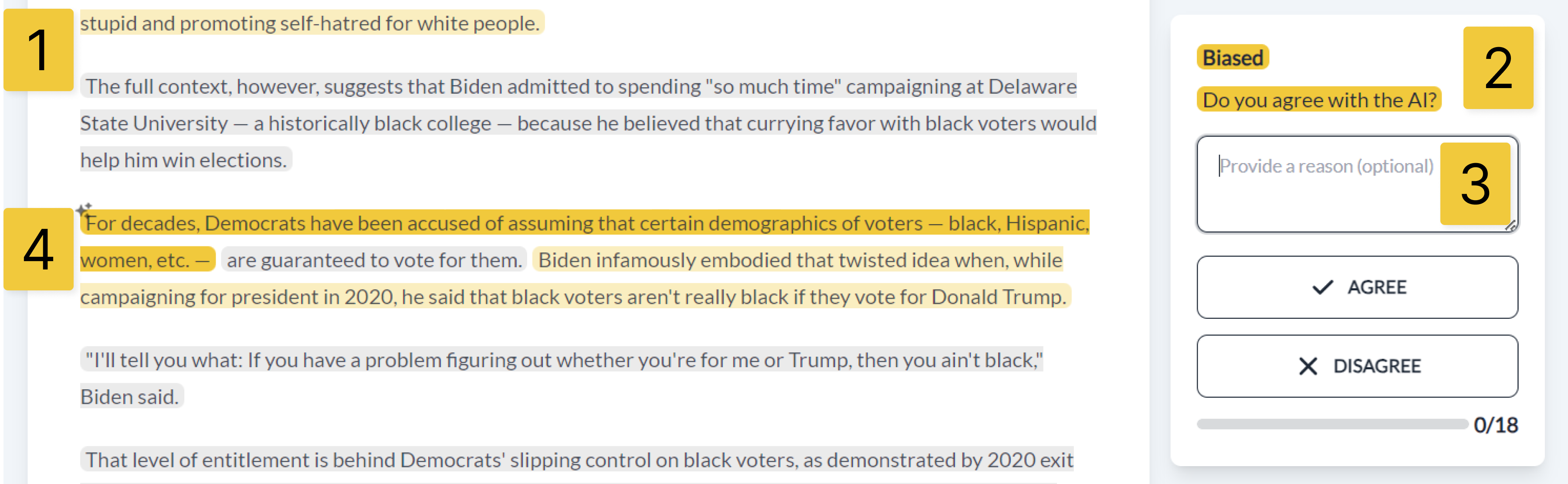}
 \caption{The classifier shows the highlights in yellow (biased) and grey (not biased) on NewsUnfold. The feedback module on the right allows readers to agree or disagree and leave optional feedback. The \textit{Sparkles} draw attention to controversial sentences or sentences that need more feedback. \Cref{table:key-elements} explains the elements with yellow numbers.}
 %\Description{The figure shows sentences from a news article and the feedback module to its right. The feedback mechanism uses the classifier to highlight biased sentences in yellow and not biased sentences in grey. Readers can agree or disagree with this classification through the feedback module on the right. The feedback module states, "biased. Do you agree with the AI?" (highlighted in yellow). Below is a free-text field for optional written feedback, an "agree" button, and a "disagree" button. The sentence in the middle is highlighted more strongly and has Sparkles at its beginning.}
 \label{fig:feedbackSparkles}
\end{figure*}

The landing page states NewsUnfold's mission: encouraging bias-aware reading and collecting feedback to refine bias detection to mitigate its negative effects.
To further motivate contributions, it emphasizes the value of individual users' feedback in enhancing bias-detection capabilities.
Clicking a call-to-action button guides users to the \textit{Article Overview Page} (\Cref{fig:articlesOverview}).
As a preliminary stage, this page displays 12 static articles spanning nine subjects, balanced by the bias amount and political orientation.
Different articles enable readers to compare the amount of bias in one article.
%added
Selecting an article directs users to NewsUnfold's \textit{Article Reading Page}, which integrates the bias highlights and feedback mechanism.
\Cref{table:key-elements} outlines its essential components.
%#4
The sparkles highlight controversial sentences or sentences that received the least feedback to enable balanced feedback collection (\inlinegraphics{figures/icons/4.png} in \Cref{fig:feedbackSparkles}).
From the \textit{Article Overview Page} (\Cref{fig:articlesOverview}), users can additionally initiate a tutorial (\inlinegraphics{figures/icons/7.png} in \Cref{fig:articlesOverview}) guiding them through the bias highlights (\inlinegraphics{figures/icons/1.png} in \Cref{fig:feedbackSparkles}), the feedback mechanism (\inlinegraphics{figures/icons/2.png} in \Cref{fig:feedbackSparkles}), and concluding with a pointer to the UX survey (\inlinegraphics{figures/icons/5.png} in \Cref{fig:recommended}).
After each article, we show three recommended articles (\inlinegraphics{figures/icons/6.png} in \Cref{fig:recommended}).

\subsection{Study Design} \label{sec:studydesign}
Our primary objectives for testing NewsUnfold in a real-world setting are:
\begin{enumerate}
  \item \textbf{Engagement:} Measure the amount of voluntary feedback from readers without monetary incentives.%, measured by single decisions made via the feedback mechanism.
  \item \textbf{Data Quality:} Assessing quality of feedback.%, based on IAA Krippendorff's alpha.
  \item \textbf{Classifier:} Investigating classifier performance when integrating feedback-generated labels.
  \item \textbf{User Experience:} Evaluating user experience and perception of NewsUnfold, focusing on bias highlights (\inlinegraphics{figures/icons/1.png} in \Cref{fig:feedbackSparkles}) and feedback (\inlinegraphics{figures/icons/2.png} in \Cref{fig:feedbackSparkles}) for a user-centered design approach.
\end{enumerate}

During the study, readers can freely explore the platform, select articles, decide to provide anonymous feedback, and choose to participate in the UX survey.
Unlike the preliminary study, participants are not sourced from crowdworking platforms but reached via LinkedIn, Instagram, and university boards.
The outreach briefly introduces NewsUnfold with a link to its landing page.
Readers are informed of feedback data collection beforehand.

To understand the readers' experiences, a voluntary UX survey (\inlinegraphics{figures/icons/5.png} in \Cref{fig:recommended}) is available after reading an article.\footnote{The survey consists of 9 questions: two scales and eight optional open-ended queries.\Cref{sec:appendix:ux} contains a detailed breakdown of the survey and its results.} 
In this study, we prioritize identifying UX issues among readers to boost participation and feedback efficiency, focusing on UX-oriented data collection over comprehensive quantitative analysis.
%We compare the feedback to findings from the preliminary study (\Cref{sec:feedback}).
To obtain user analytics, we use \textit{Umami}\footnote{\url{https://umami.is}}, a privacy-centric tool logging the number of clicks, unique visitors, country, language settings, device types, most-visited pages, and the number of tutorial initiations while keeping the anonymity of visitors.

\subsection{Dataset Creation and Evaluation}\label{sec:NUDA}
NewsUnfold collects anonymous feedback on bias highlights on 12 articles with 357 sentences at the sentence level.
The data is stored on university servers.
Articles cover topics consistent with the baseline dataset (e.g., gender equality, black lives matter, and climate change \cite{spindeNeuralMediaBias2021}), represent different political slants, and are balanced regarding bias strengths.
NewsUnfold uses a repeated-labeling method \cite{Sheng_Zhang_2019}, employing a majority-vote system with a minimum of five votes per sentence to establish sentence labels.
%#2
The labels are stored in the same structure as BABE \cite{spindeNeuralMediaBias2021} to enable the merging of the two datasets.
We apply a spam detection method by \citet{spammer} to filter out unreliable annotations.
We calculate a score between 0 and 1 for each annotator and eliminate annotators in the 0.05th percentile.
We assess the quality of the resulting dataset, similar to \Cref{sec:feedback}, using the IAA metric Krippendorff's $\alpha$ and manual analysis.

As HITL systems center around iteratively improving machine performance through user input, we evaluate the integration of feedback data into classifier training.
The training process adopts hyperparameter configurations from \citet{spindeNeuralMediaBias2021} with a pre-trained model from Hugging Face.\footnote{\url{https://huggingface.co/mediabiasgroup/DA-RoBERTa-BABE}}
We train and evaluate the model with data from NUDA added to the 3700 BABE sentences and compare it against the baseline classifier \cite{spindeNeuralMediaBias2021} using the F1-Score \cite{F1score}.

\section{Results} \label{sec:results}
From March 4th to March 11th (2023), NewsUnfold had 187 unique visitors.
158 read articles, 33 (20.89\%) provided sentence feedback, and eight offered 25 additional reasons for feedback, mainly on sentences perceived as biased (84\%) but highlighted as not biased (80\%).
45 (28.48\%) completed the tutorial, and 13 (6.9\%) the UX survey.
Geographically, 61\% were from Germany, 25\% from Japan, 6\% from the United States.
Language-wise, 45\% preferred English, 42\% preferred German.
Notably, 52\% accessed via mobile, highlighting mobile optimization's importance.\footnote{We detail all statistics on \url{https://doi.org/10.5281/zenodo.8344891}.}

\begin{figure*}[ht]
 \includegraphics[width=0.8\textwidth]{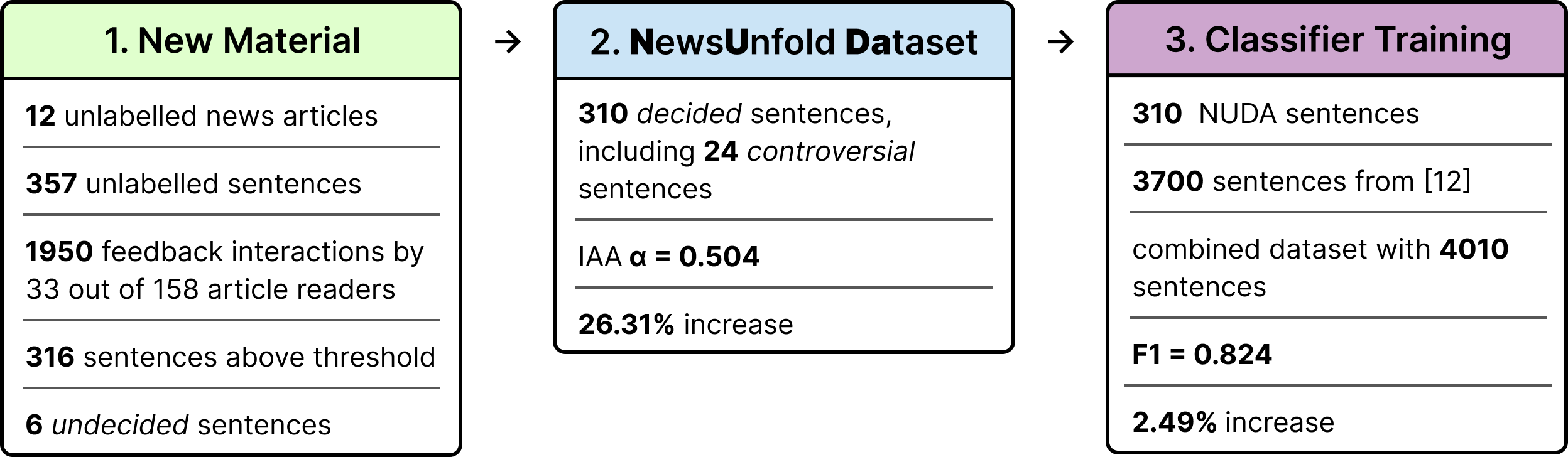}
 \centering
 \caption{The NewsUnfold dataset creation process. First, readers incorporate and annotate new material. Second, all decided sentences are collected into NUDA, and their IAA is calculated. Third, NUDA is added to BABE for classifier training.}
 %\Description{Three steps from new material to the NUDA dataset to the final classifier training are displayed. The new material (first tile) consists of 12 news articles with 357 sentences. 1950 feedback interactions by 33 visitors resulted in 316 sentences above the threshold. NUDA (second tile) has 310 decided sentences, including 24 controversial sentences, with an IAA of 0.504, an increase of 26.31\%. Classifier training (third tile) then combines the 310 NUDA sentences with the 3700 BABE sentences and achieves an F1 score of 0.824, a 2.49\%.}
 \label{fig:dataset}
\end{figure*}

The 357 sentences collectively received 1997 individual annotations, representing either agreement or disagreement with the presented classifier outcome.
We identify two spammers within the 5\% spammer score range and remove 47 annotations, leaving 1950 valid annotations in the dataset.
316 sentences attain a label through the repeated-labeling method.
A sentence is categorized as \textbf{decided} if there is a majority, \textbf{controversial} if the biased-to-unbiased feedback ratio lies between 40-60\%\footnote{A sentence can be \textbf{decided} and \textbf{controversial} at the same time.}, and \textbf{undecided} if the ratio stands at an exact 50\% as listed in \Cref{fig:dataset}.
310 \textbf{decided} sentences spanning nine topics form NUDA.\footnote{For the full dataset, see \url{https://doi.org/10.5281/zenodo.8344891}}

\subsection{Data Quality} \label{sec:results:dataquality}
\begin{figure}[ht]
 \includegraphics[width=0.47\textwidth]{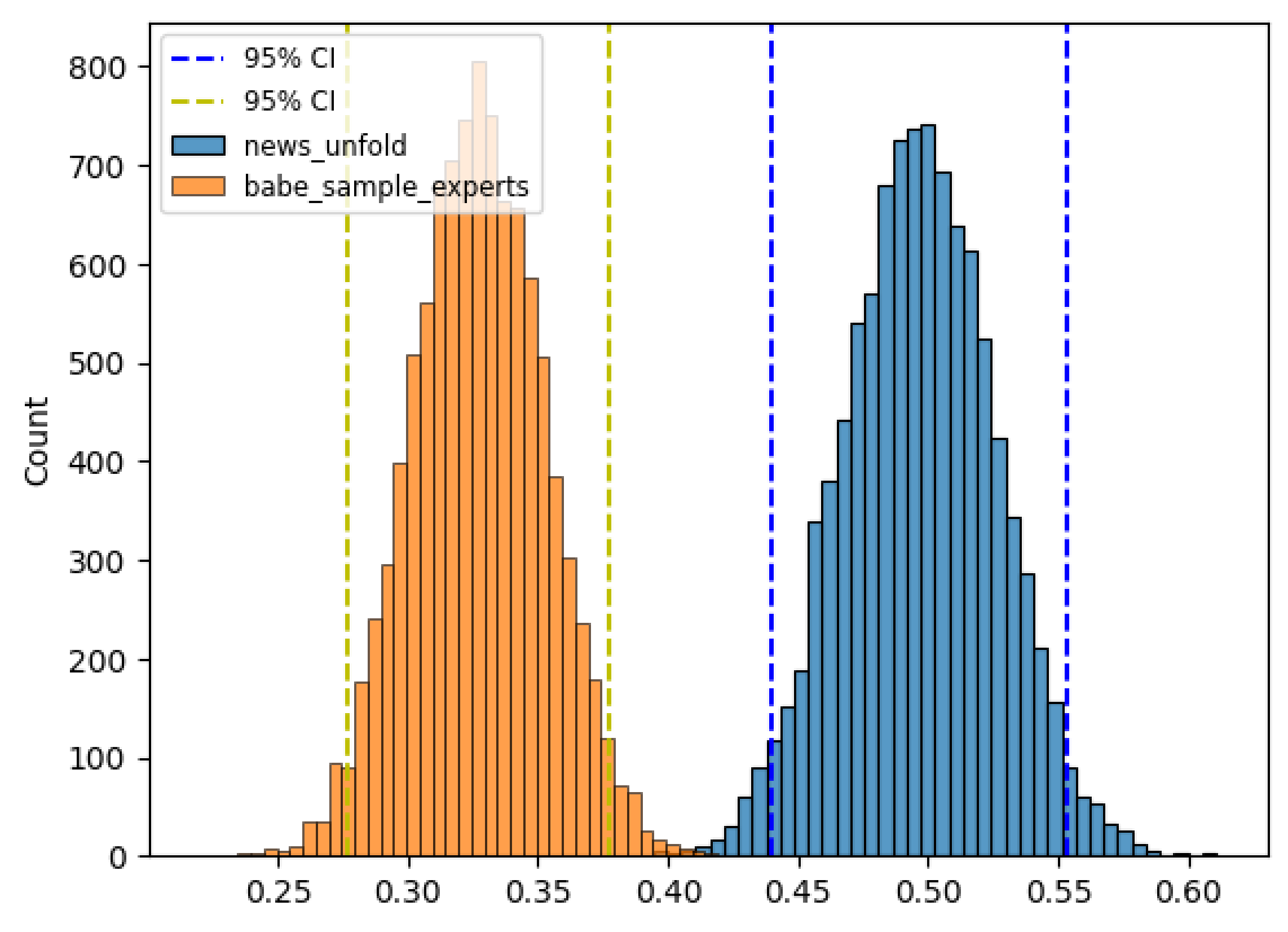}
 \caption{Comparison of the expert-generated dataset with the NUDA dataset. The non-overlapping confidence intervals indicate a significant increase.}
 %\Description{The bootstrapped confidence intervals of the expert-generated dataset (orange) with the NUDA dataset (blue) do not overlap.}
 \label{fig:significance}
\end{figure}

To evaluate if NewsUnfold increases data quality, we calculate the Inter-Annotator agreement score Krippendrorff's $\alpha$.
The NUDA dataset achieves a Krippendorff's $\alpha$ of .504.
The 26.31\% increase in IAA compared to the baseline's IAA of .399 \cite{spindeNeuralMediaBias2021} is statistically significant, as demonstrated in \Cref{fig:significance} by the non-overlapping bootstrapped confidence intervals.
To demonstrate that the IAA does not merely increase with the sample size but through higher data quality, we take 100 randomly sized dataset samples ($n = 10$ to $n = 1950$), calculate the IAA for each, and employ a regression model.

The model's explanatory power ($R^2=.009$, $R^2{\scriptscriptstyle adjusted}=-.002$) suggests a negligible linear relationship between sample size and the F1 score \Cref{tab:OLSResults}.
This implies that the model does not explain the variance in F1 scores when accounting for the increase in data points.
Moreover, the F-statistic of .8424 ($p = 0.361$) does not provide evidence to reject the null hypothesis that there is no linear relationship between sample size and F1 score ($x{\scriptscriptstyle1}=.000004$, $SD=.000004$, $t=-.918$, $CI[.00001, .000004]$).
Therefore, we conclude that the collected data is reliable, and increases in quantity do not necessarily translate into increased data quality.
Further, we conducted a manual evaluation by annotating 310 sentences and comparing these expert annotations against the labels provided by NUDA.
The comparison yielded an agreement of 90.97\% across 282 labels, with a disagreement of 9.03\% over 28 labels.
Specifically, the experts identified 25 sentences as biased, which NUDA had not, whereas only three sentences deemed biased by the experts were classified as unbiased by NUDA.
A closer examination of the disagreeing labels revealed that the primary source of discrepancy was sentences containing direct quotes.
When we removed 69 sentences predominantly consisting of direct quotes, the agreement increased to 95.44\% on 230 labels, with the disagreement rate dropping to 4.56\% on 11 labels.
Of these, ten sentences experts labeled as biased were not labeled as biased by NUDA, and one sentences experts labeled as biased was labeled not biased by NUDA.
This high agreement rate suggests that NewsUnfold can gather high-quality annotations and labels.

\subsection{Classifier Performance}
After merging NUDA with the BABE dataset, the average F1 score (5-fold cross-validation) is .824 (\Cref{tab:F1Scores}), showing a 2.49\% improvement over the BABE baseline \cite{spindeNeuralMediaBias2021}.
While this may not constitute a substantial improvement, it is a positive increment towards the anticipated direction.
%added
%However, the increase might be due to more data collected during the test run and not exclusively to HITL. %added
We conduct five 5-fold cross-validations with different distributions to control for potential biases in the F1-Score due to imbalanced dataset distribution.
Folds and repetitions show only marginal differences with a variance of .000022, suggesting that the data quality provides reliable results.
%This demonstrates that the dataset collected is of high enough quality to yield reliable results, positioning NewsUnfold as a feasible method for improving media bias datasets.

\begin{table}[]
\centering
\begin{tabular}{l|c|c}
\toprule
Dataset & Sentences & F1-Score (\%)\\\midrule
BABE & 3700 & .804 ± .014  \\
NUDA and BABE & 4010 & .824 ± .017 \\\bottomrule
\end{tabular}
\caption{Comparison of classifiers trained on BABE alone versus BABE combined with NewsUnfold Dataset.}
\label{tab:F1Scores} 
\end{table}

\subsection{User Experience Survey Results}
Thirteen participants took part in the UX survey.
They express positive feelings about the platform and bias highlights (\Cref{ux:highlights}).
The platform's ease of use receives a high rating of 8.46 on a 10-point scale, indicating a user-friendly design, affirmed by participants' descriptions of the interface as intuitive and concise.
While almost all users state a positive effect on reading more critically, some raise concerns about highlight calibration, their ineffectiveness with unbiased articles, and bias introduced by direct quotes in news articles.

Participants exhibit varied opinions when providing feedback, most enjoying it, some undecided, and one finding it work-like (\Cref{ux:givingfeedback}).
For those interested in giving feedback, the survey indicates an easy process.

One participant mentioned that skipping the tutorial leads to confusion.
Thus, one could consider making the tutorial mandatory in future iterations.
%The positive evaluation of the ease of use facilitates high retention rates and engagement.
%Further, we expect that the self-reported heightened bias awareness positively correlates with data quality.
In conclusion, we expect that the ease of use facilitates higher retention rates and engagement while the self-reported heightened media bias awareness positively correlates with data quality.

\section{Discussion}\label{sec:Discussion}
\subsection{Feedback Mechanisms Study}
Although feedback was optional, monetary incentives and a structured study setting prompted participants to share opinions on all highlights, raising questions about engagement in settings without such incentives.
Initially, we assumed the \textit{Comparison} method reduced anchoring bias and increased critical thinking.
However, F1 scores between \textit{Highlights} and \textit{Comparison} disprove this.
Both F1 and IAA scores were expectantly low as media bias perception is highly subjective, and comparable approaches report similar scores \cite{spindeMBICMediaBias2021, PhrasingBiasHube2019, recasens2013a}.
%added
\textit{Comparison} was less efficient than \textit{Control}, possibly due to managing two questions simultaneously.
Interestingly, the \textit{Highlights} method led to longer engagement times, indicating a mix of focused attention and prolonged article interaction, possibly enhancing contextual critical thinking.

%#16
Further, \Cref{tab:FeedbackArticles} highlights issues with the training data for BABE,\footnote{Students primarily annotated the data.} treated as the ground truth.
Discrepancies between expert labels suggest that BABE may not be entirely accurate, especially since the dataset often misclassifies subtly biased sentences as "not biased".
However, achieving complete accuracy in bias classification may be unattainable due to the subjective nature of bias and the misleading concept of a single, absolute ground truth \cite{xuNoteBiasComplete2024}.

\subsection{NewsUnfold}
NewsUnfold showcases how the feedback mechanism gathers bias annotations in a news-reading environment.
The system increases IAA by 26.31\%, achieves high agreement with expert labels, and corrects misclassifications through feedback.
For example, this sentence was initially deemed non-biased but corrected to biased:
\begin{center}
\textit{"That level of entitlement is behind Democrats' slipping control on black voters, as demonstrated by 2020 exit polls showing that, for example, just 79\% of black men voted for Biden, a percentage that has been dropping since 2012."}
\end{center}
Despite having a lower label count than BABE, the feedback dataset demonstrates greater agreement with expert labels.
%#15%
Furthermore, statistical analysis indicates that the improvement in IAA cannot be attributed solely to the annotation count \Cref{sec:results:dataquality}.
Although the increase in the dataset size likely drives the rise in F1-Score, the data has been shown to be reliable.
This suggests that readers using the feedback mechanism (\inlinegraphics{figures/icons/2.png} in \Cref{fig:feedbackSparkles}) offer a reliable alternative to costly expert annotators, facilitating the collection of more extensive data sets.
The scarcity of high-quality media bias datasets highlights the need to integrate feedback mechanisms on NewsUnfold or other digital and social platforms.
Likewise, other classifiers, such as misinformation classifiers, can use similar mechanisms to gather data and improve accuracy while augmenting the cognitive abilities of readers \cite{pennycookAccuracyPromptsAre2022a, spindeHowWeRaise2022}.

Limited written feedback through the feedback mechanism (\inlinegraphics{figures/icons/3.png} in \Cref{fig:feedbackSparkles}) might be due to typing disruptions during reading.
Most feedback highlighted false negatives, indicating it is simpler to spot bias than explain its absence.
Currently, nuances are not well-captured by the binary feedback in the first iteration, as all design decisions are a trade-off between an effortless process that drives engagement and more complex labeling.
While more friction can foster deeper thinking, we decided on a simple, binary feedback version (\inlinegraphics{figures/icons/1.png} in \Cref{fig:feedbackSparkles}).
%added
Scales similar to \citet{karmakharm-etal-2019-journalist} could turn binary feedback into a spectrum and include multiple scales for other biases.% \cite{spindeIntroducingMediaBias2022}. % and their perception by including labeled scales.

Although direct quotes can exhibit bias, they do not inherently impact neutrality \cite{recasens2013a}.
In our dataset, we observe significant disagreements regarding quotes (\Cref{sec:results:dataquality}), indicating confusion among readers regarding their interpretation.
Therefore, future iterations should incorporate different visual cues for quotes and may consider excluding them from the bias indication and training dataset.

Expanding the data collection phase could have enlarged the dataset but potentially bear design flaws.
Hence, we decide to follow a user-centric design approach with a short collection phase to allow for quick iterations of feedback while showcasing data quality capabilities early on.
While the RoBERTa model fine-tuned with BABE was used, NewsUnfold could have tested other models.
However, RoBERTa performed superior in a previous study \cite{spindeNeuralMediaBias2021}.

%#9
A common challenge in projects that rely on community contributions is keeping volunteers motivated over time \cite{crowdMotivation}.
With NewsUnfold, we aim to increase motivation by highlighting bias in a news reading application, offering a reason for people to use the platform that goes beyond annotation.
%#10
The project targets reader groups similar to those interested in AllSides and GroundNews, which have demonstrated the viability of such business concepts.
For testing and iterative feedback, we opted for a binary approach, feasible with our resources at the time, predicated on the assumption that feedback would primarily come from individuals valuing unbiased information.
In later versions, NewsUnfold will incorporate insights from a recent literature review on media bias detection and mitigation \cite{xuNoteBiasComplete2024}.
The authors suggest accounting for cultural and group backgrounds in label creation and output generation.
By adapting its output to readers' backgrounds, NewsUnfold could extend its appeal beyond those specifically seeking unbiased information.
Gamification elements or unlocking additional content through giving feedback could further increase motivation \cite{CrowdGamification}.

The use cases of the feedback mechanisms extend beyond NewsUnfold.
Any digital and social media that includes text can apply the feedback mechanism to raise readers' awareness and collect feedback.
NewsUnfold, as an application, integration, or browser plug-in, could offer an alternative to traditional news platforms.
Incorporating feedback mechanisms with customizable classifiers, such as those for detecting misinformation, stereotypes, emotional language, generative content, or opinions, could allow users to analyze the content they consume in greater detail.
Simultaneously, they contribute to a community dataset with an open-source purpose, which has shown potential in other applications \cite{cooperPredictingProteinStructures2010}.
We believe that by offering something useful, the feedback mechanism on NewsUnfold can gather valuable information in the long run, even if readers do not interact with it daily.

%added for R1: example for malicious groups + methods to prevent their impact
While systems like NewsUnfold can help understand bias and language, educate readers, and foster critical reading, we must closely monitor data quality and include readers' backgrounds while meeting data protection standards.
Bias in the reader base or attacks from malicious groups, for example, any politically extreme group interested in shifting the classifier according to their ideology, could lead to a self-enforcing loop that inserts bias and skews classifier results towards a specific perspective, potentially harming minorities.
To avoid deliberate attacks, we include spammer detection before training.
In the future, we will monitor feedback beyond F1 scores and IAA as they only capture the agreement between raters, which is a standard measure in the media bias domain but does not fully indicate the quality of annotations.
They help us set a baseline to check how bias detection systems handle human feedback, backed up by the manual analysis and NewsUnfold's ease of use.% and the increased performance of NUDA.
Other possibilities include a soft labeling approach \cite{fornaciariBlackWhiteLeveraging2021}, adding adversarial examples into the training data, 
\cite{goyalSurveyAdversarialDefenses2023}, employing a HITL approach where experts try to break the model \cite{wallaceTrickMeIf2019}, identifying and correcting perturbations \cite{goyalSurveyAdversarialDefenses2023}, or using a more complex probabilistic model for label generation \cite{Law2011}.

Making the system and process transparent is critical to avoid misuse.
Given the potential impact of skewed or misclassified bias highlights on reader perceptions, the system must communicate the impossibility of achieving absolute accuracy.
Hence, the landing page informs readers about the possible inaccuracy to impart a clear understanding of classification limitations and ask for readers' help.%call to action to give feedback
We believe that even with the classification improvements of large language models such as GPT, assessing human perception of bias will always be crucial, and feedback mechanisms to assess such perception are becoming more critical for developing and constantly evaluating fair AI.

\subsection{Limitations}
Our team's Western education might add bias.
%#12
Similarly, the Prolific study and the recruitment for NewsUnfold via LinkedIn might skew results due to the presence of more academic participants.
%#5
The age range and education (\Cref{fig:v1:educationAge}) of participants in \Cref{sec:feedback} suggests a bias towards the digitally accustomed and educated, additional to a left slant (\Cref{fig:v1:polMean}).
%#11
Although both studies involve relatively small samples, the results are nevertheless significant.
Future research should examine larger and more diverse samples to evaluate how varying backgrounds and political orientations influence feedback behavior and quality.
%However, proofing data quality doesn't rely on sample size.
%#7
%as real life as possible
%to test
We implemented the feedback mechanism on the NewsUnfold platform to test if readers would give feedback in an environment as close as possible to a real news aggregator.
Hence, we decided against a demographic survey to collect annotator data as it might negatively impact readers' experience.
The more open study setting \citep[compare][]{spindeMBICMediaBias2021, spindeTASSYTextAnnotation2021}, with users exploring NewsUnfold freely, complicates the identification of factors affecting data quality.
While the goal is to gather data from diverse readers, NewsUnfold currently controls for geographical diversity.
Unlike the US pre-study, NewsUnfold had significant participation from Japan and Germany.
However, readers' backgrounds and quality control tasks must be implemented in later iterations, for example, by implementing user accounts.
Their data can be used to improve models \cite{Law2011} and fair classifiers \cite{cabitzaPerspectivistTurnGround2023} accounting for backgrounds and protecting minorities or underprivileged groups.

%%#8 oof
We did not collect demographic data or ask participants which device they used to view NewsUnfold in the UX study.
Later studies need to control for situations, experiences, attention, and perceptions of bias, which could diverge depending on personal backgrounds and the device used.

\subsection{Future Work} \label{sec:futurework}
We plan to develop NewsUnfold into a standalone website with constantly updated content.

Simultaneously, we aim to evaluate different feedback mechanisms for media bias classifiers and to extend our design's application beyond NewsUnfold.
We plan to implement and test the feedback tool (\inlinegraphics{figures/icons/2.png} in \Cref{fig:feedbackSparkles}) as a browser plugin\footnote{A future experiment will determine if a plugin ensures quality annotations. However, its platform integration expands reach and diversity, making it our preferred choice.} and social media integration.
%#3
The value of the feedback mechanism lies in its adaptability across different platforms, using visual cues to enhance datasets for various bias types.
Social media websites can add similar mechanisms as extensions highlighting biased language to make users more aware of potential biases and their influence \cite{spindeHowWeRaise2022}.
Our next phase involves testing its integration in social media environments like X.

We will monitor the impact on user behavior in the long term and explore gamification and designs to increase engagement \cite{gamificationHITL2022}.
Also, we will assess varied bias indicators, such as credibility cues \cite{NudgeCredBhuiyan_2021}, which have shown to be effective in similar studies \cite{yaqubEffectsCredibilityIndicators2020, kenningSupportingCredibilityAssessment2018}, but need real-world validation.
We further plan to add labels for subtypes of bias \cite{spindeIntroducingMediaBias2022}.
Advanced models like LLaMA \cite{touvronLLaMAOpenEfficient2023}, BLOOM \cite{workshopBLOOM176BParameterOpenAccess2023}, and GPT-4 \cite{openaiGPT4TechnicalReport2023} may offer additional explanations on bias highlights (\inlinegraphics{figures/icons/1.png} in \Cref{fig:feedbackSparkles}).
When both models and data collection improve, it facilitates finding and comparing different outlets' coverage of topics and views of the general population.
Simultaneously, controlling for personal backgrounds \cite{biasStrategiesGoerling2013} could assist journalists and researchers in studying and understanding media bias, as well as its formation and expression over time.% added

\section{Conclusion}\label{sec:conclusion}
We present NewsUnfold, a HITL news-reading application that visually highlights media bias for data collection.
It augments an existing dataset via a previously evaluated feedback mechanism, improving classifier performance and surpassing the baseline IAA while integrating a UX study.
NewsUnfold showcases the potential for diverse data collection in evolving linguistic contexts while considering human factors.

\section{Acknowledgments} 
This work was supported by the Hanns-Seidel Foundation (\url{https:// www.hss.de/}), the German Academic Exchange Service (DAAD) (\url{https://www.daad.de/de/}), the Bavarian State Ministry for Digital Affairs in the project XR Hub (Grant A5-3822-2-16), and partially supported by JST CREST Grant JPMJCR20D3 Japan.
None of the funders played any role in the study design or publication-related decisions.
ChatGPT was used for proofreading.

%todo put references here, right order: main content, appendix, ethics, acks, references
\bibliography{sample-base}

\appendix
%\section{Appendix}
\label{sec:app}

%here demographics
\section{Feedback Mechanism Study Texts}
\subsection{Data Processing Agreement} \label{sec:appendix:data}
\textbf{Who are we and how do we use the data we collect from you through this survey?}
This research study is being conducted by the Media Bias Research Group.
We are a group of researchers from various disciplines with the goal of developing systems and data sets to uncover media bias or unbalanced coverage in articles.
This study is anonymous.
That means that we will not record any information about you that could identify you personally or be associated with you.
On the basis of the collected data, we aim to publish scientific papers on presentations of articles that help to detect biased language, but these publications do not allow any inference to you as an individual.
Once the study is published, the anonymized data might be made available in a public data repository.
Your rights to access, change, or move your information are limited insofar as the data may no longer be modified after the data has been published in anonymized form.
The reason for this is that we need to manage your information in specific ways for the research to be reliable and accurate.
Once anonymized, we will not be able to delete your data.
The study itself is not hosted on Prolific, but on a dedicated external server.
Once the survey is complete, you will be shown a unique code that you can enter in the Prolific form.
Participation in this study is voluntary.
You may choose not to participate and you may withdraw at any time during the study without any penalty to you.
If you have any questions about the study or study procedures, you may contact the Media Bias Research Group, info@media-bias-research.org.
\begin{itemize}
\item I agree to the processing of my personal data in accordance with the information provided herein.(Checkbox)
\end{itemize}
\subsection{Demographic Survey} \label{sec:demoSurv}
\begin{enumerate}
\item What gender do you identify with? (Female, Male, Other, Prefer not to say)
\item What is your age? (Input field for number)
\item What is the highest level of education you have completed?
(8th grade, Some high school, High school graduate, Vocational or technical school, Some college, Associate degree, Bachelor’s degree, Graduate work, Ph.D., I prefer not to say)
\item What is the level of your English proficiency? (Proficient, Independent, Basic)
\item Do you consider yourself to be liberal, conservative, or somewhere in between? Please slide to record your response. (Very liberal to Very conservative, -10 to 10 point slider)
\item How often on average do you check the news? (Never, Very rarely, Several times per month, Several times per week, Every day, Several times per day)
\end{enumerate}
\subsection{Info on Media Bias}
Before you can start we will now provide you with a few examples that should help you to understand possible media bias instances better. For each example, a sentence with a biased word (blue colored) is shown first followed by its impartial representation (green colored).
Please note that bias is different from negative sentiment. Bias is ambiguous and subtle, it can be positive, negative, or not even have a particular sentiment but it still can imply or intensify the opinion/emotion.

\textbf{Subjective Intensifiers:}

Schnabel himself did the fantastic reproductions of Basquiat’s work.

Schnabel himself did the accurate reproductions of Basquiat’s work.

\textbf{Strong labels:}

'The people want the Truth!': Trump gloats over the loss of American media jobs.

'The people want the Truth!': Trump tweets over the loss of American media jobs.

\textbf{One-sided terms:}

Concerned Women for America’s major areas of political activity have consisted of opposition to gay causes, pro-life

law...

Concerned Women for America’s major areas of political activity have consisted of opposition to gay causes,

anti-abortion law...

\subsection{Attention Check on Bias}
\textbf{How is bias connected to sentiment?}
Based on the information that was provided to you earlier, please select the correct option.
\begin{itemize} 
\item Bias is the same as negative sentiment.
\item Bias can be both positive, negative or even not have particular sentiment. (correct answer)
\item Bias is the same as positive sentiment.
\item Bias is not connected to sentiment at all.
\end{itemize}

\subsection{Trust Check}
\textbf{Can we trust your data for scientific research?}
For example, if you failed to pay attention to some questions, please answer 'No'. Please answer honestly, you will receive full payment regardless of your answer.
Please select one option. (Yes, you can trust my data for scientific research. No, you may not want to trust my data for scientific research.)

\section{Detailed UX Survey Results for NewsUnfold} \label{sec:appendix:ux}
\subsubsection{How did you like NewsUnfold? (10 responses)} \label{ux:like}
Six participants expressed a strongly positive sentiment, stating, for instance, that they found it innovative. Two expressed that the bias detection might need some calibration, one found it "okay", and one remained unsure.
\subsubsection{Ease of Use: How easy was NewsUnfold to use? (13 responses)}\label{ux:easeofuse}
Participants were asked to rate the ease of use of NewsUnfold on a 10-point scale, with 10 indicating high ease of use. The average rating for ease of use was 8.46, with a median score of 9, implying that users found NewsUnfold user-friendly and intuitive.
\subsubsection{How did NewsUnfold impact your reading? (12 responses)}\label{ux:readingimpact}
6 participants stated it made them read more carefully, critically, and slowly instead of skimming. Two stated bias was easier to recognize because they had to think twice. One said they did more active thinking about what bias is. One didn't feel much impact in unbiased articles. One wanted it in all of their browsing. One said it made them argue with the AI instead of skimming the article.
\subsubsection{How did you feel about giving feedback on the sentences? (11 responses)}\label{ux:givingfeedback}
Three participants found it easy to give feedback, while two reported it felt either difficult because it disrupted their reading flow or because it felt like a chore.
Two participants felt unsure, with one skipping the tutorial. Two reported only doing it when they would have more time. One stated only to give feedback when disagreeing with the classification. One participant appreciated sharing their reasoning in the free-text field \inlinegraphics{figures/icons/3.png}.
\subsubsection{How do you feel about the highlights in the text? (10 responses)}\label{ux:highlights}
Nine participants liked the highlights and found them helpful, one calling it their favorite part. However, one participant found it distracting and raised concerns about highlighting quotes as biased.
\subsubsection{Net Promoter Score (NPS): How likely would you recommend NewsUnfold to a friend, family, or colleague? (13 responses)}\label{ux:NPS}
The calculated mean NPS was 6.23. This score indicates participants were neutral to slightly in favor of recommending NewsUnfold.
\subsubsection{How do you like the User Interface of NewsUnfold? (11 responses)}\label{ux:ui}
9 participants found it easy and clean, with one stating the "look is sleek and appropriate for a modern website". One participant experienced a bug using Firefox on mobile and described it as a "bit sluggish." One participant found the UI "a bit bland."% concise, easy and accessible, sleek and appropriate for a modern website, and smooth and impressive.
\subsubsection{What irritated you? Did you encounter any problems? (10 responses)}\label{ux:irritated}
Bugs and irritations included the character limit in the free-text field \inlinegraphics{figures/icons/3.png}, the multiple steps in the feedback window on mobile devices, overlays blocking the text on Firefox mobile, out-of-line tooltips, and encountering jumping buttons. One person expressed a slight annoyance in instances they disagreed with the classifier. One person was confused because they skipped the tutorial. Two didn't encounter problems.
\subsubsection{Anything else you want to share with us? (1 response)}\label{ux:additional}
One participant suggested that it might be interesting to add a note indicating that direct quotes are more likely to be biased and may not necessarily reflect the opinions of the authors.

\section{Material Bias and Demographics of Feedback Mechanism Study}

\begin{table*}[h!]
\centering
\renewcommand{\arraystretch}{0.8} % Reduces the height of rows
\setlength\tabcolsep{3pt} % Further reducing the space between columns
\scriptsize % Setting a smaller font size
\begin{tabular}{l|cccccccccccccc}
\toprule
\rotatebox{90}{Dep. Variable} & \rotatebox{90}{R-squared} & \rotatebox{90}{Adj. R-squared} & \rotatebox{90}{F-statistic} & \rotatebox{90}{Prob (F-statistic)} & \rotatebox{90}{Log-Likelihood} & \rotatebox{90}{No. Observations} & \rotatebox{90}{AIC} & \rotatebox{90}{BIC} & \rotatebox{90}{Df Residuals} & \rotatebox{90}{Df Model} & \rotatebox{90}{Covariance Type} & \rotatebox{90}{Const coef} & \rotatebox{90}{x1 coef} \\
\midrule
y & 0.009 & -0.002 & 0.8424 & 0.361 & 240.50 & 100 & -477.0 & -471.8 & 98 & 1 & nonrobust & 0.0553 & .000004 \\
\bottomrule
\end{tabular}
\caption{OLS Regression Results for F1 Score of the NewsUnfold Feedback}
\label{tab:OLSResults}
\end{table*}

%\section{Demographics Feedback Mechanism Study} 
\begin{figure}[h!]
 \includegraphics[width=0.4\textwidth]{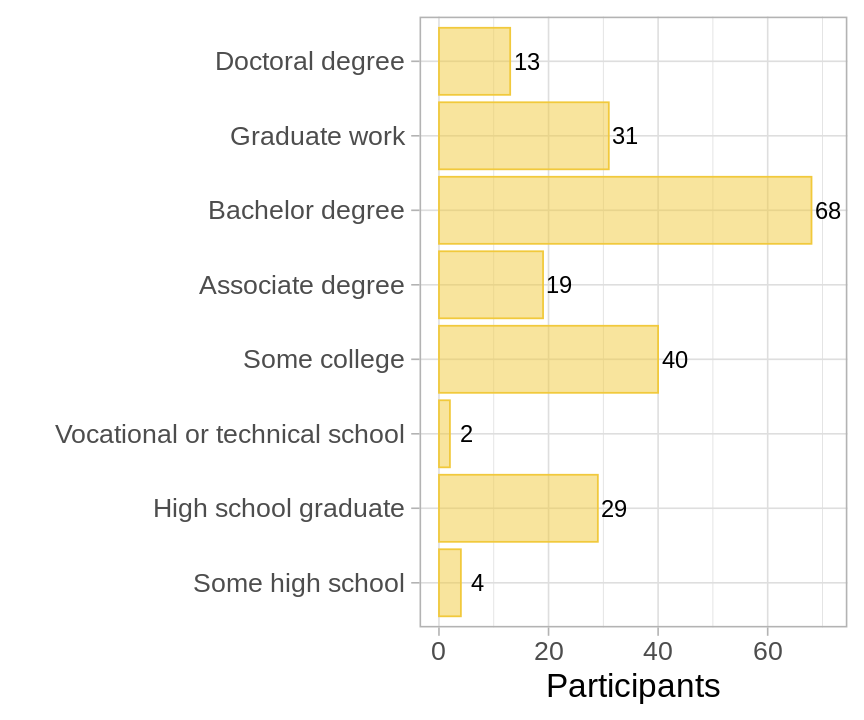}
 \caption{Education of participants in the feedback mechanism study.}
 \label{fig:v1:education}
\end{figure}

\begin{figure}[h!]
 \includegraphics[width=0.4\textwidth]{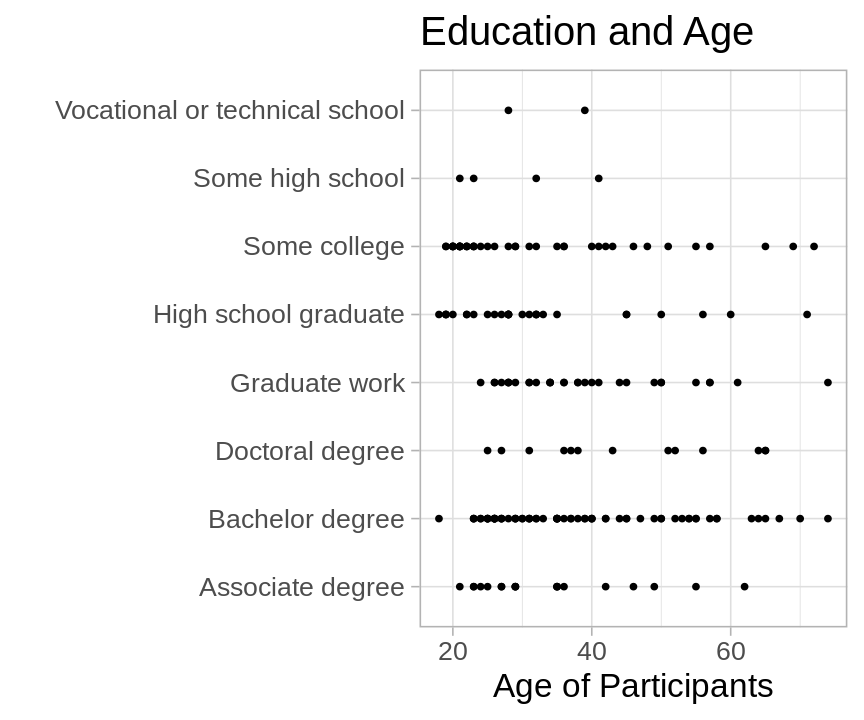}
 \caption{Education of participants mapped to their age in the feedback mechanism study.}
 \label{fig:v1:educationAge}
\end{figure}

\begin{figure}[h!]
 \includegraphics[width=0.4\textwidth]{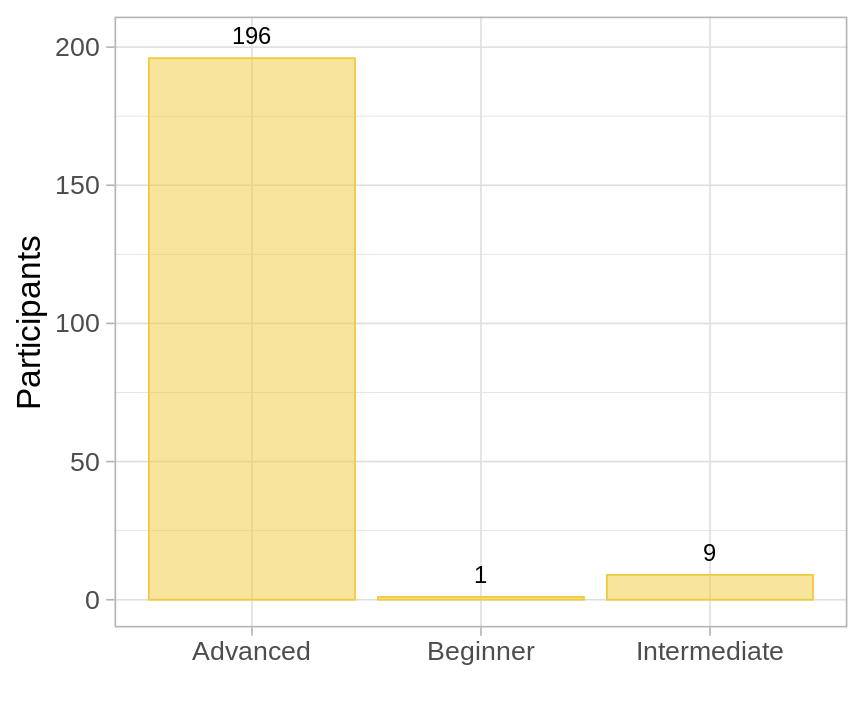}
 \caption{English proficiency of participants in the feedback mechanism study.}
 \label{fig:v1:english}
\end{figure}

\begin{figure}[h!]
 \includegraphics[width=0.4\textwidth]{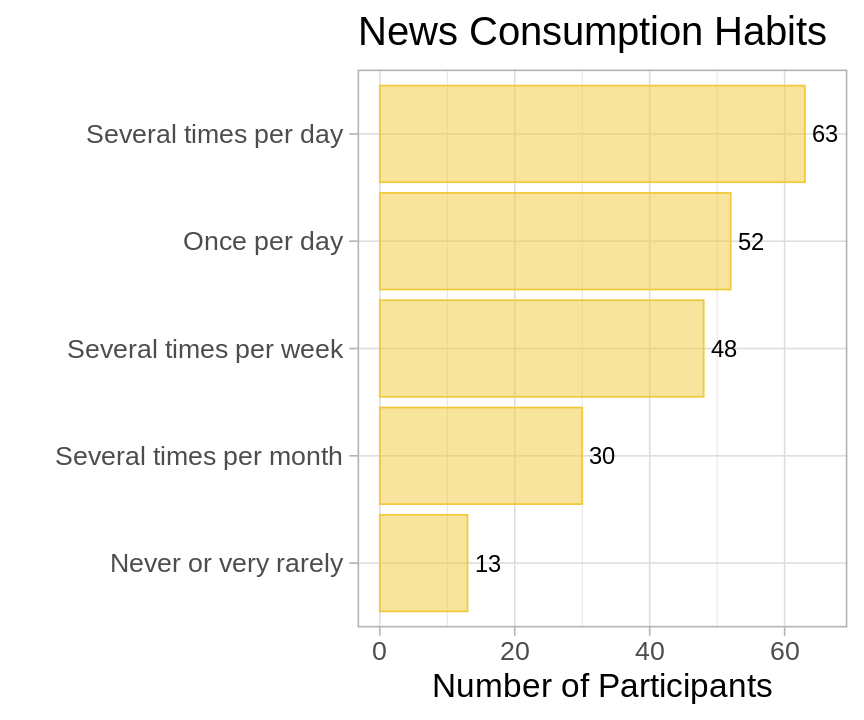}
 \caption{News consumption habits of participants in the feedback mechanism study.}
 \label{fig:v1:news}
\end{figure}

\begin{figure}[h!]
 \includegraphics[width=0.4\textwidth]{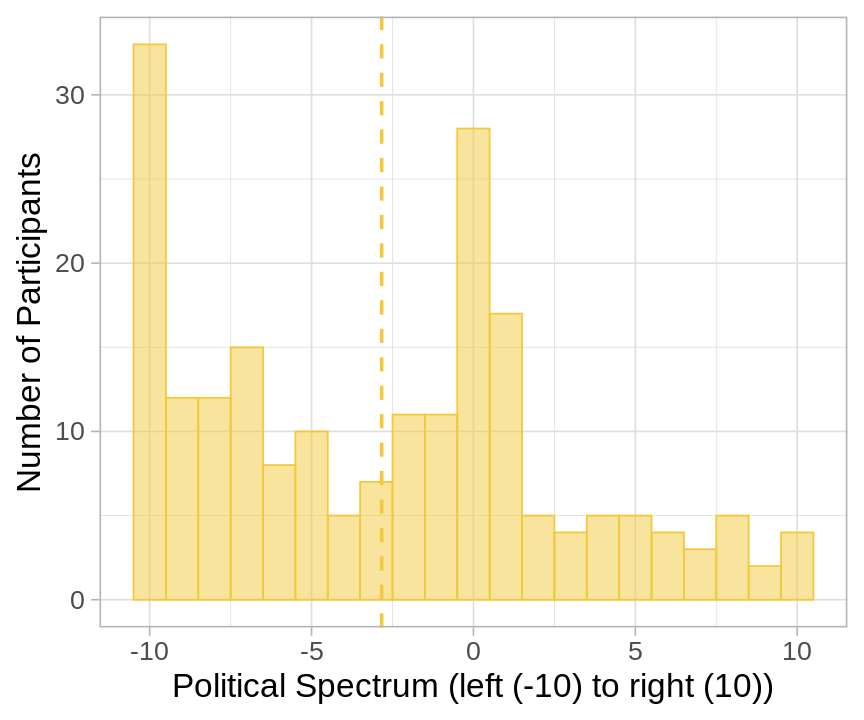}
 \caption{Political orientation of participants in the feedback mechanism study.}
 \label{fig:v1:pol}
\end{figure}

\begin{figure}[h!]
 \includegraphics[width=0.4\textwidth]{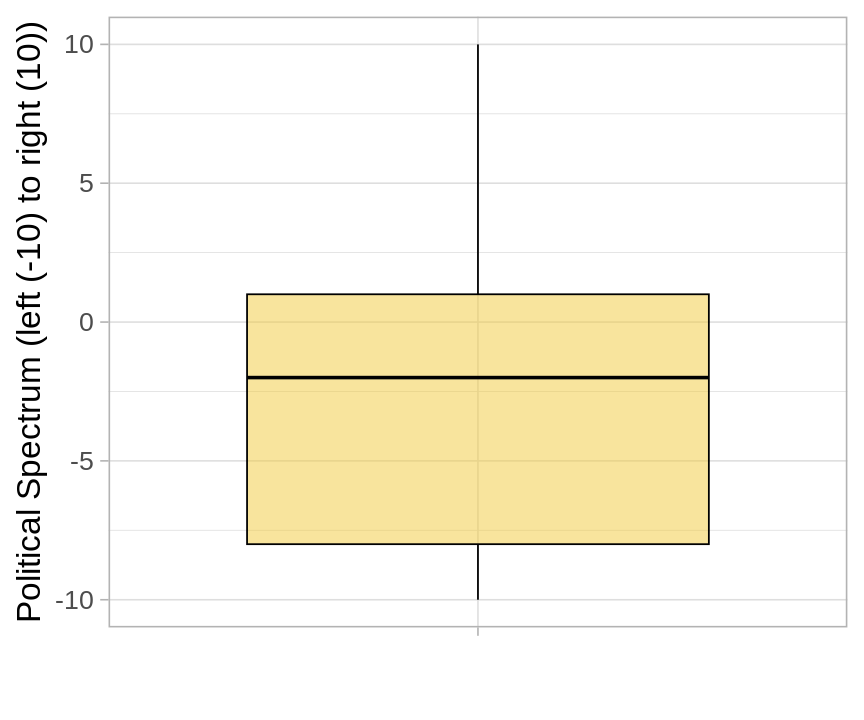}
 \caption{Average political orientation of participants in the feedback mechanism study.}
 \label{fig:v1:polMean}
\end{figure}

\begin{table*}[h!]
\centering
\begin{tabular}{l|c|c|c|c}
\toprule
 & Experts biased & Experts not biased & Classifier biased & Classifier not biased\\\midrule
Left article & 16 & 21 & 8 & 29 \\
Right article & 24 & 21 & 12 & 33 \\\bottomrule
\end{tabular}
\caption{Bias rating of sentences in feedback mechanism study articles by classifier and experts.}
\label{tab:FeedbackArticles} 
\end{table*}

\twocolumn[\section{Additional Screenshots}] \label{sec:appendix:pics}
\begin{figure}[h!]
 \includegraphics[width=\textwidth]{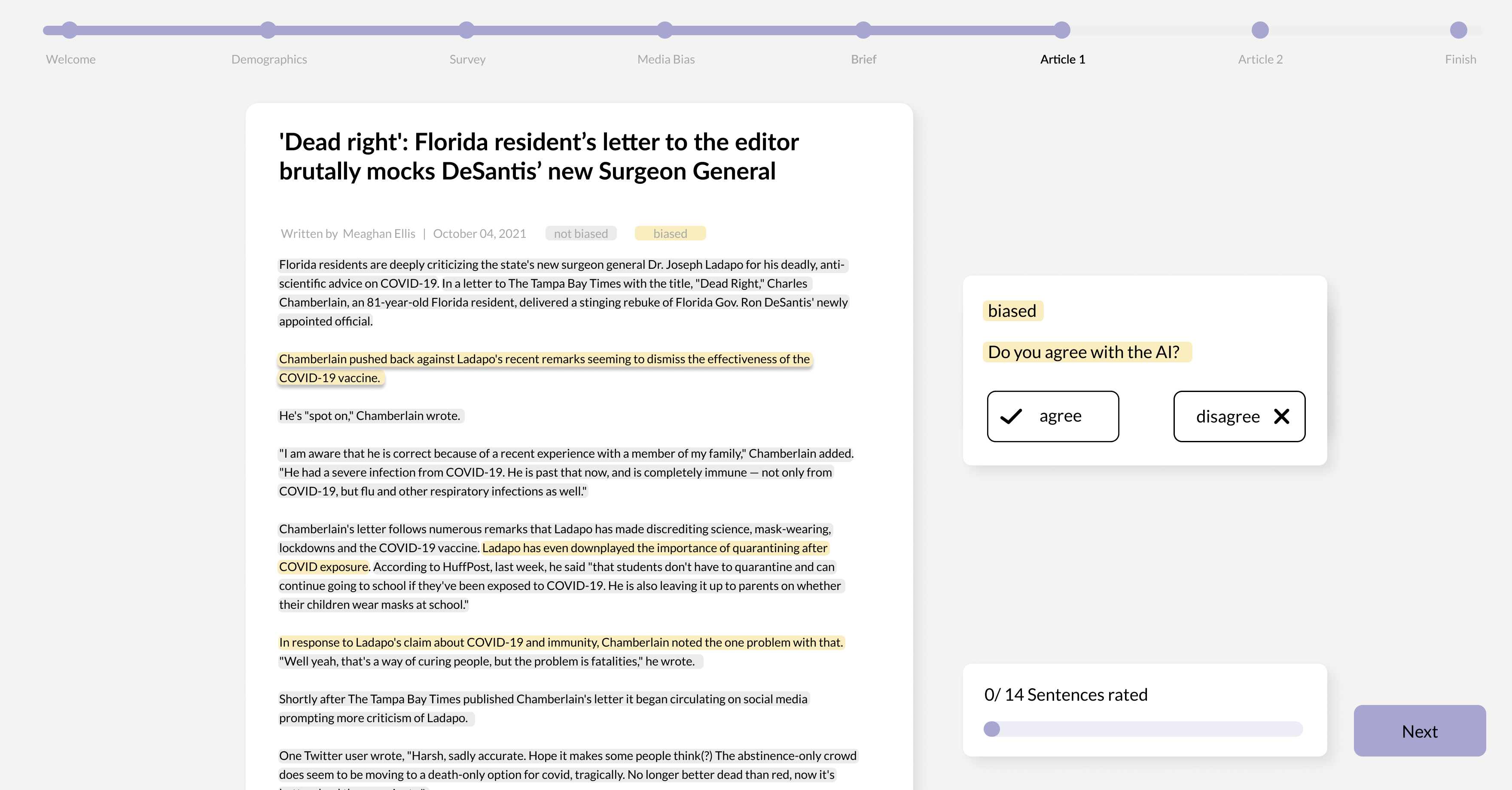}
 \caption{Screenshot of the highlight mechanism on the study platform for the preliminary feedback mechanism study.}
 %\Description{A screenshot showing the study design with the article text and highlights in the middle and the highlights feedback mechanism on the right side. On top is a progression bar that leads participants through the study. In the lower right corner, a bar fills up with each feedbacked sentence.}
 \label{fig:FeedbackHighlights}
\end{figure}

\begin{figure}[h!]
 \includegraphics[width=\textwidth]{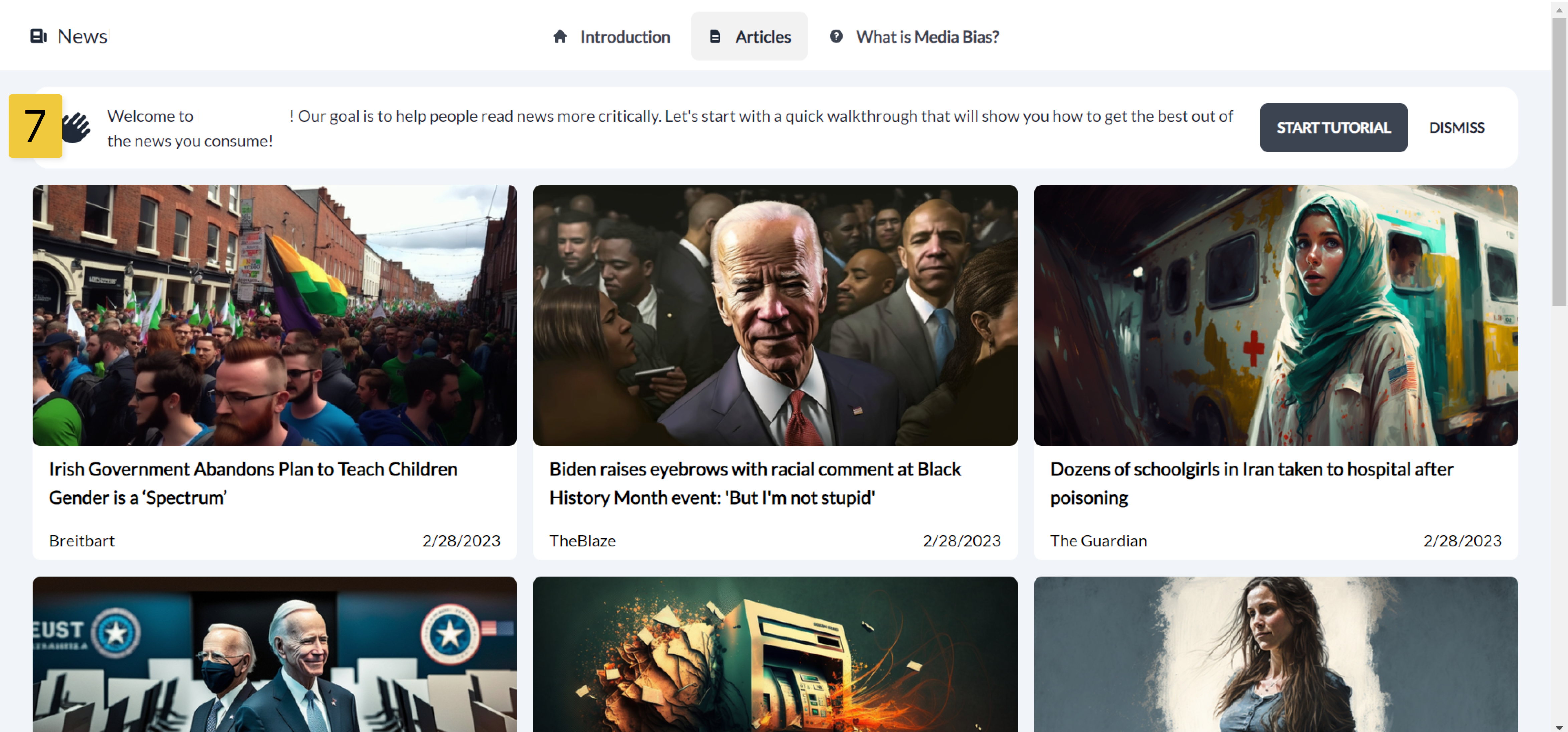}
 \caption{Screenshot of Articles Overview Page with the option to start the tutorial. \Cref{table:key-elements} explains the elements with yellow numbers.}
 %\Description{Articles Overview with tiles for multiple articles, including a picture, headline, date, and outlet, and the option to start the tutorial over a button.}
 \label{fig:articlesOverview}
\end{figure}

\begin{figure*}[h!]
 \includegraphics[width=\textwidth]{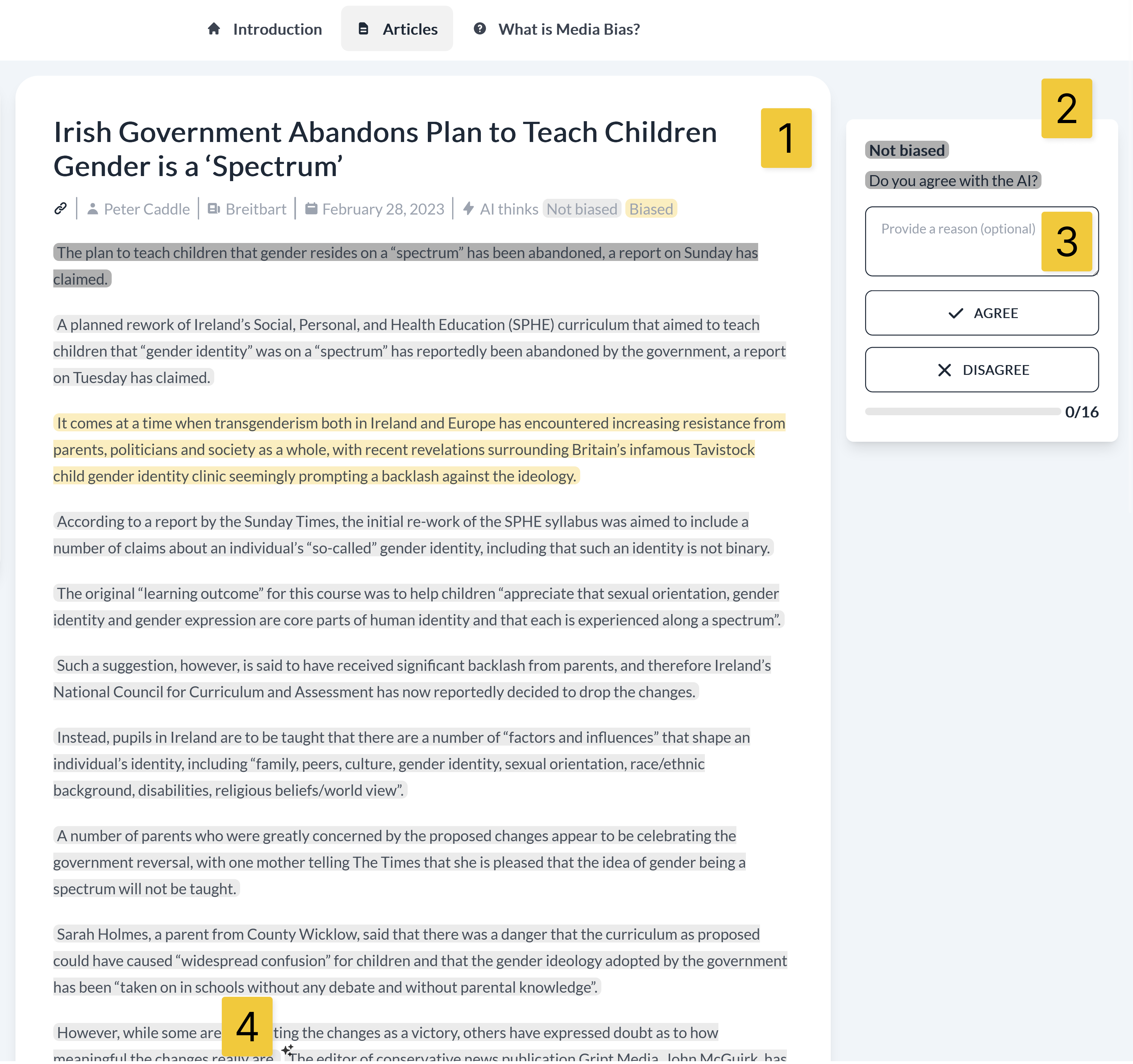}
 \caption{NewsUnfold Article View. \Cref{table:key-elements} explains the elements with yellow numbers.}
 %\Description{The article with highlights in the middle and the feedback mechanism on the right. In addition to the previous figures, this figure displays the headline, article metadata, and the website's navigation bar ("Introduction," "Articles" (highlighted), "What is Media Bias."}
 \label{fig:NewsUnfold}
\end{figure*}

\begin{figure*}[h!]
 \includegraphics[width=\textwidth]{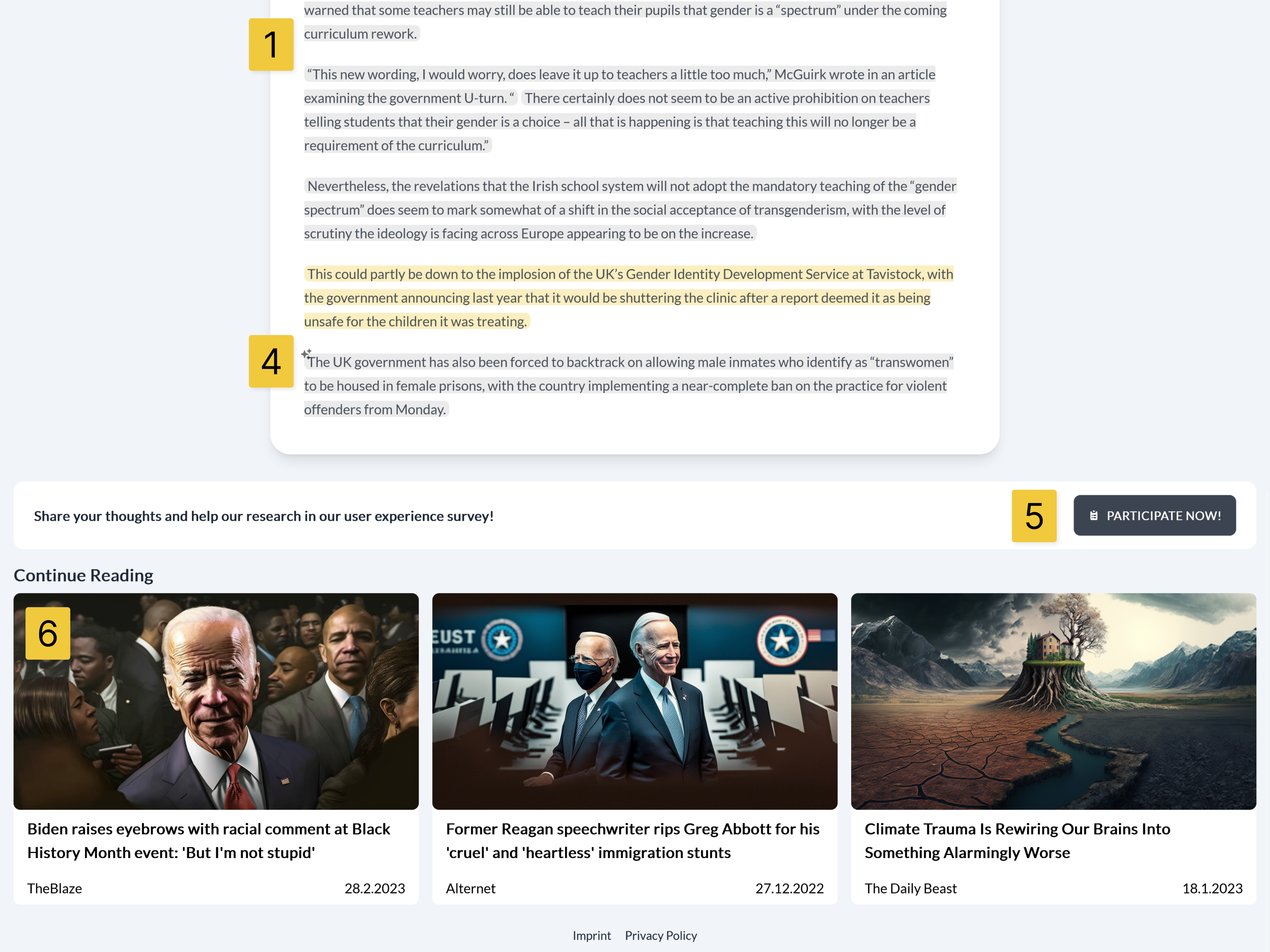}
 \caption{Screenshot of recommended articles and button to the UX survey. \Cref{table:key-elements} explains the elements with yellow numbers.}
 %\Description{Three recommended articles with pictures, headlines, date, and outlet shown beneath the article. Above it, a button takes readers to the UX survey.}
 \label{fig:recommended}
\end{figure*}

\begin{figure*}[ht!]
\centering
 \includegraphics[width=0.8\textwidth]{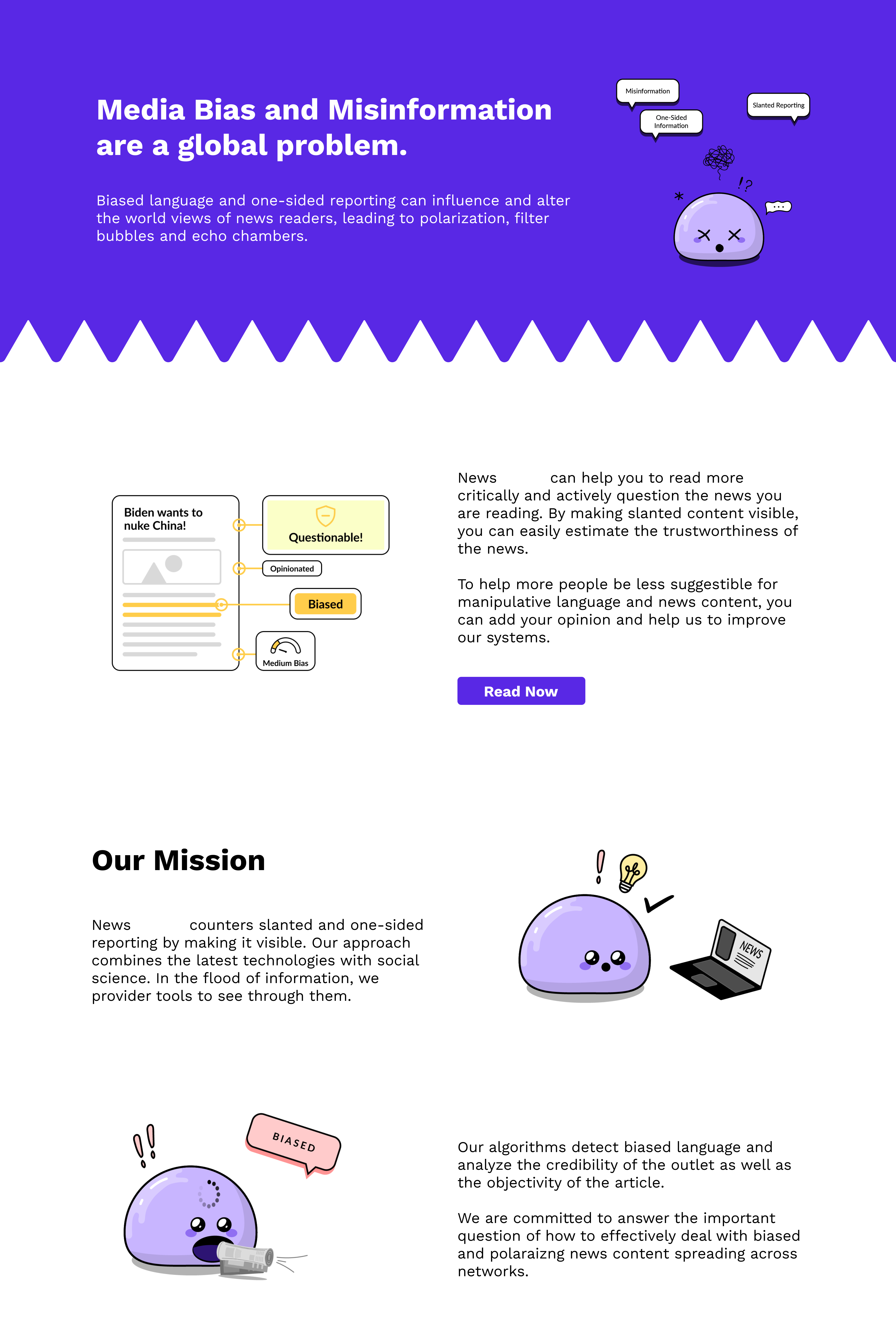}
 \caption{Screenshot of the landing page with an introduction to NewsUnfold's goal.}
 \label{fig:landingPage}
\end{figure*}

\end{document}